\documentclass[12pt]{article}
\pdfoutput=1
\usepackage{putex}
\usepackage{comment}
\usepackage{graphicx}
\usepackage{caption}
\usepackage{amsmath}
\usepackage[usenames,dvipsnames,table]{xcolor}
\usepackage{array}
\usepackage{subcaption}
\usepackage{epstopdf}
\usepackage{enumerate}
\usepackage{cite}
\usepackage{tensor}
\usepackage{slashed}
\usepackage[export]{adjustbox}
\usepackage[aligntableaux=center]{ytableau}
\usepackage{dsfont}
\usepackage[utf8]{inputenc}
\usepackage[
      colorlinks=true,
      linkcolor=blue,
      urlcolor=blue,
      filecolor=black,
      citecolor=red,
      ]{hyperref}
\newcommand{\RNum}[1]{\uppercase\expandafter{\romannumeral #1\relax}}

\newcommand {\be} {\begin {equation}}
\newcommand {\ee} {\end {equation}}

\newcommand {\bes} {\begin {equation*}}
\newcommand {\ees} {\end {equation*}}

\newcommand{\beq}{\begin{equation}}
\newcommand{\eeq}{\end{equation}}

\def\ie{\begin{equation}\begin{aligned}}
\def\fe{\end{aligned}\end{equation}}

\numberwithin{equation}{section}

\def\<{\langle}
\def\>{\rangle}

\begin{document}

%\preprint{PUPT-}

\institution{PU}{Department of Physics, Princeton University, Princeton, NJ 08544, USA}

\title{Fermions in AdS and Gross-Neveu BCFT} 

\authors{Simone Giombi, Elizabeth Helfenberger, and Himanshu Khanchandani}

\abstract{We study the boundary critical behavior of conformal field theories of interacting fermions in the Gross-Neveu universality class. 
By a Weyl transformation, the problem can be studied by placing the CFT in an anti de Sitter space background. 
After reviewing some aspects of free fermion theories in AdS, we use 
both large $N$ methods and the epsilon expansion near 2 and 4 dimensions to study the conformal boundary conditions in the Gross-Neveu CFT. 
At large $N$ 
and general dimension $d$, we find three distinct boundary conformal phases. Near four dimensions, where the CFT is described by the Wilson-Fisher 
fixed point of the Gross-Neveu-Yukawa model, two of these phases correspond respectively to the choice of Neumann or Dirichlet boundary condition 
on the scalar field, while the third one corresponds to the case where the bulk scalar field acquires a classical expectation value. 
One may flow between these boundary critical points by suitable relevant boundary deformations. We compute the AdS free energy on each of them, 
and verify that its value is consistent with the boundary version of the F-theorem. We also compute some of the BCFT observables 
in these theories, including bulk two-point functions of scalar and fermions, and four-point functions of boundary fermions. }

\date{}
\maketitle

\tableofcontents

\section{Introduction and summary}
It is well-known that a given CFT can have multiple conformally invariant boundary conditions. These correspond to different boundary critical points of the same bulk CFT, and they may be connected by Renormalization Group (RG) flows triggered by operators localized on the boundary. The critical behavior of a CFT in the presence of a boundary may be described using the language of boundary conformal field theory (BCFT), which is defined by the spectrum of local operators on the boundary and their OPE coefficients, in addition to the usual bulk CFT data and new bulk-boundary OPE data (see for instance \cite{McAvity:1995zd, Billo:2016cpy} for an introduction to the subject of BCFT). Perhaps the simplest example is the CFT of a free massless scalar field. In the presence of a boundary, there are two conformally invariant boundary conditions: one may impose Neumann or Dirichlet boundary conditions. This leads to two inequivalent BFCTs. One can flow from the Neumann to the Dirichlet theory by adding a boundary mass term, which is a relevant deformation for the Neumann boundary condition. The situation is much richer in interacting theories, see for instance \cite{Diehl:1996kd, McAvity:1995zd, Liendo:2012hy, Carmi:2018qzm, Giombi:2020rmc} for the canonical case of the interacting scalar CFT with $O(N)$ invariant interactions. 

The purpose of this paper is to study theories with fermions in the presence of a boundary. Previous works on fermionic theories with conformal boundary conditions include for instance \cite{McAvity:1993ue, Herzog:2017xha, Herzog:2018lqz, Carmi:2018qzm}. To be more specific, we consider the Gross-Neveu (GN) model in dimensions $2 < d < 4$, which is a theory of $N$ Dirac fermions with an $U(N)$ invariant interaction \cite{PhysRevD.10.3235}
\begin{equation} \label{ActionGN}
S = - \int d^d x \sqrt{g} \left( \bar{\Psi}_i \gamma\cdot \nabla \Psi^i +  \frac{g}{2} (\bar{\Psi}_i \Psi^i)^2 \right).
\end{equation}
The coupling $g$ is dimensionless in two dimensions, and the model has a perturbative UV fixed point in $d = 2 + \epsilon$. At large $N$, the critical point of the  model may be described by introducing an auxiliary Hubbard-Stratonvich field and dropping the quadratic term $\sim \sigma^2$ which becomes irrelevant in the critical limit (see for instance \cite{Moshe:2003xn, Giombi:2016ejx} for reviews). This yields the following action that can be used to develop the $1/N$ expansion of the large $N$ CFT \footnote{Throughout this paper, we are always going to assume that the bulk theory is at its critical point.}
\begin{equation} \label{ActionGNLarge}
S = - \int d^d x \sqrt{g} \left( \bar{\Psi}_i \gamma \cdot \nabla \Psi^i + \sigma \bar{\Psi}_i \Psi^i \right).
\end{equation}
This leads to a unitary conformal field theory in the dimension range $2 < d < 4$ and the $1/N$ perturbation theory for this model is well studied. The main goal of this paper is to study the behavior of this theory in the presence of a boundary. 

\begin{figure} 
\centering
\includegraphics[scale=0.9]{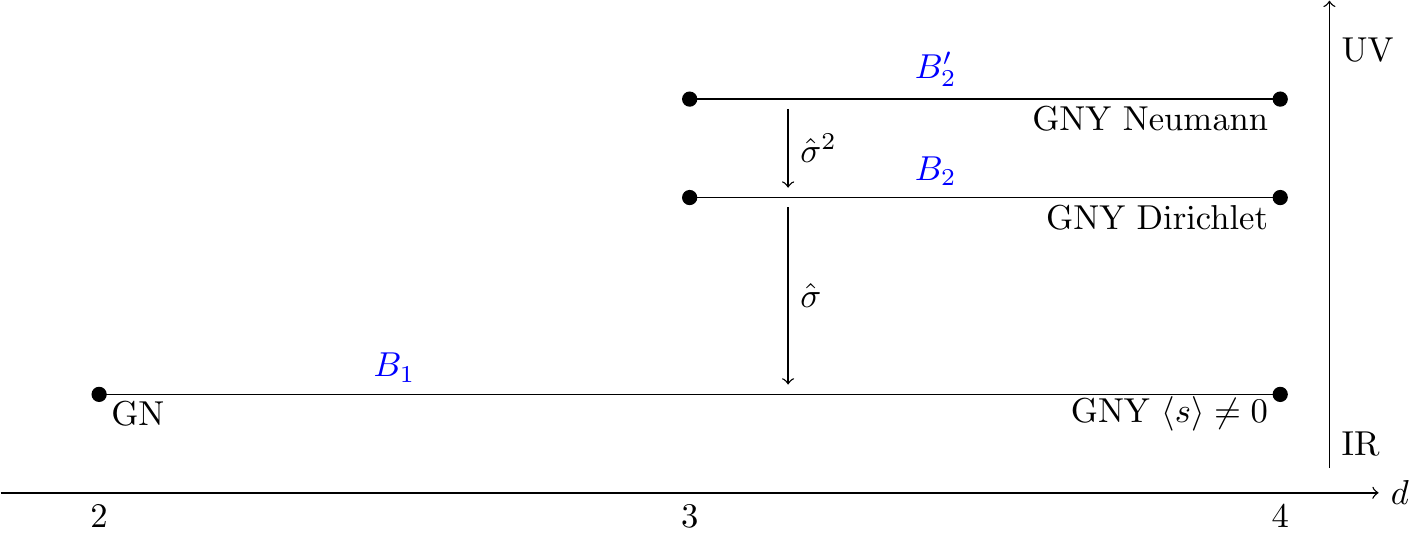}
\caption{Various boundary phases for the large $N$ model and their description in terms of Gross-Neveu and Gross-Neveu Yukawa model near $d = 2$ and $4$ respectively. The figure is not to scale, and the top two phases can only be distinguished at subleading order in $1/N$.}
\label{FigureBoundaryPhases}
\end{figure}

A CFT in flat space with a flat (or spherical) boundary is Weyl equivalent to the CFT in AdS \cite{Paulos:2016fap, Carmi:2018qzm, Herzog:2019bom, Herzog:2020lel, Giombi:2020rmc}.\footnote{See also \cite{CALLAN1990366, Aharony:2010ay, Aharony:2015hix, Hogervorst:2021spa, Antunes:2021abs} for related studies of QFT and CFT in AdS.} Under this map, the boundary is mapped to the asymptotic boundary of the hyperbolic space. This enables us to directly apply the results from the extensive AdS/CFT literature about fermions in AdS \cite{Henneaux:1998ch, Mueck:1998iz, Arutyunov:1998ve, Mueck:1999efk, Henningson:1998cd, Kawano:1999au, Basu:2006ti, Aharony:2010ay, Aros:2011iz, Faller:2017hyt, Nishida:2018opl} to the problem of BCFT. %The correlation functions of the boundary operators can be calculated by using Witten diagrams in AdS. 
Throughout this paper, we use this AdS description to do calculations. At leading order at large $N$, we obtain the following results that summarize the boundary critical behavior of the Gross-Neveu CFT: If we impose that the boundary spectrum satisfies unitarity bounds, then in dimensions $2 < d < 3$, there is a single boundary conformal phase characterized by the leading fermion operator with scaling dimension $\hat{\Delta}_{\left( 1/2 \right)} = \hat{\Delta} = d - 3/2+O(1/N)$ in its boundary spectrum. We call this phase $B_1$. However as we go above three dimensions, in  $3 \leq d < 4$, in addition to the above, there is another possible unitary phase, which has $\hat{\Delta}_{\left( 1/2 \right)} = \hat{\Delta} = d - 5/2+O(1/N)$. At subleading order in $1/N$, we find that this actually splits into two distinct phases, which we call $B_2$ and $B_2'$. They have different bulk two-point functions for the fluctuations of the $\sigma$ field around the saddle point (which is the same for $B_2$ and $B_2'$ cases, corresponding to the same large $N$ boundary fermion dimension), in particular yielding a different scaling dimension $\hat{\Delta}_{\left( 0 \right)}$ for the leading boundary scalar induced by the $\sigma$ field. Let us note that, in all of the boundary conformal phases, we find that the bulk-boundary OPE of the bulk field $\sigma$ includes a scalar operator of dimension $d$, which corresponds to the displacement operator (the presence of such operator is required by conformal symmetry in any BCFT). To summarize, near four dimensions, there are a total of three boundary critical points of the model. See Figure \ref{FigureBoundaryPhases} where we summarize various phases and the RG flows between them. 

As shown in \cite{Zinn-Justin:1991ksq}, there is another description of the GN model in terms of the IR fixed point of the Gross-Neveu-Yukawa (GNY) model which has $N$ Dirac fermions and a single scalar
\begin{equation}
S = \int d^d x  \left( \frac{(\partial_{\mu} s)^2}{2} - \left( \bar{\Psi}_i \gamma\cdot \nabla \Psi^i + g_1 s \bar{\Psi}_i \Psi^i\right) + \frac{g_2}{24} s^4   \right).
\end{equation} 
This model is weakly coupled near $d = 4$ and one may develop a perturbation theory in $\epsilon$ in $d = 4 - \epsilon$, where one finds an IR fixed point (see for instance \cite{Fei:2016sgs} for a more extensive review and several results on the CFT data at this fixed point). The field $s$ is essentially identified with $\sigma$ in (\ref{ActionGNLarge}), up to rescaling by the coupling constant $g_1$. To be consistent with the results in the large $N$ description, we expect to find three boundary phases in the GNY description as well. Indeed as we will show in section \ref{Sec:EpsExp}, the phases $B_2$ and $B_2'$ correspond to doing perturbation theory around Dirichlet or Neumann boundary condition for the scalar $s$ respectively, while $B_1$ corresponds to having a classical vev for the field $s$. In this sense, near four dimensions, $B_2$ and $B_2'$ are analogous to the so-called ordinary and special transition in $O(N)$ scalar BCFT, while $B_1$ is analogous to extraordinary transition (see e.g. \cite{Diehl:1996kd, Liendo:2012hy, Giombi:2020rmc}). For easy reference, in Table \ref{TableBoundaryPhasesDimensions} we report the dimensions of the boundary operators induced by the bulk fundamental fields $\Psi$ and $\sigma$, along with the dimensions of the same operators in the $\epsilon$ expansion description, which can be seen to be precisely consistent with each other.

\begin{table}[!ht]
\centering
\begin{tabular}{|c|c|c|c|}
\hline
 & Large $N$  & GNY $d = 4 - \epsilon$ & GN $d = 2 + \epsilon$  \\
\hline
 & & & \\
$B_1$ & $\hat{\Delta}_{\left( 1/2 \right)} = d - \frac{3}{2}$ & $\hat{\Delta}_{\left( 1/2 \right)} = \frac{3}{2} + \sqrt{\frac{36}{-2 N + 3 + \sqrt{4 N ^2 + 132 N + 9}}}  + O(\epsilon)$ & $\hat{\Delta}_{\left( 1/2 \right)} = \frac{1}{2} + \frac{4 N - 3}{4 (N - 1)} \epsilon $   \\
 & & & \\
 & $\hat{\Delta}_{\left(0\right)} = d $ & $\hat{\Delta}_{\left(0\right)} = 4 + O(\epsilon)$ & $\hat{\Delta}_{\left(0\right)} = 2 + \epsilon$ \\
\hline
 & & & \\
$B_2$ & $\hat{\Delta}_{\left( 1/2 \right)} = d - \frac{5}{2}, $ & $ \hat{\Delta}_{\left( 1/2 \right)} =  \frac{3}{2} - \frac{(8 N + 7)}{4 (3 + 2 N)} \epsilon $ & - , \\
 & & & \\
 & $\hat{\Delta}_{\left(0\right)} = 2$  & $\hat{\Delta}_{\left(0\right)} = 2  - \frac{\sqrt{4 N^2 + 132 N + 9} - 2 N + 21}{12 (2 N + 3)} \epsilon$ & -  \\
\hline 
 & & & \\
 $B_2'$ & $\hat{\Delta}_{\left( 1/2 \right)} = d - \frac{5}{2},$ &  $\hat{\Delta}_{\left( 1/2 \right)} = \frac{3}{2} - \frac{(8 N + 9)}{4 (3 + 2 N)} \epsilon $ &  -   \\
 & & & \\
 & $\hat{\Delta}_{\left(0\right)} = d - 3$ & $\hat{\Delta}_{\left(0\right)} = 1  - \frac{\sqrt{4 N^2 + 132 N + 9} + 22 N + 21}{12 (2 N + 3)} \epsilon$  & -  \\
\hline
\end{tabular}
\caption{The dimensions of the leading boundary operators induced by the bulk fundamental fields $\Psi$ and $\sigma$ at large $N$ in the three boundary phases we find. We also show the corresponding results from $\epsilon$ near two and four dimensions. The phase in the first row exists for $2 < d < 4$, while the two phases in the bottom two rows only exist between $3 < d < 4$.}
\label{TableBoundaryPhasesDimensions}
\end{table}

In the free scalar BCFT, one may flow from Neumann to Dirichlet boundary conditions by turning on a boundary mass term. In the GNY model, we still expect this to be true near four dimensions, and one should be able to flow from $B_2'$ to $B_2$ by turning on $\hat{s}^2$, where $\hat{s}$ is the leading operator in the boundary operator expansion of $s$. In the large $N$ theory, the role of $s$ is played by the $\sigma$ field, hence in the large $N$ theory, the flow from $B_2'$ to $B_2$ must be driven by $\hat{\sigma}^2$ operator. Continuing the analogy with $O(N)$ scalar BCFT, to flow from ordinary to extraordinary transition there, one can turn on the analog of a ``boundary magnetic field''.  In the GNY model description this corresponds to turning on the $\hat{s}$ operator on the boundary, and in the large $N$ description to turning on $\hat{\sigma}$. So we should be able to flow from the $B_2$ to $B_1$ phase by turning on the $\hat{\sigma}$ operator at the boundary. We will see in section \ref{Sec:LargeN} that there is no relevant scalar in the boundary spectrum for the $B_1$ phase, so we expect $B_1$ to be the most stable in the RG sense, followed by $B_2$ and $B_2'$.

Following a similar proposal for bulk CFT in \cite{Giombi:2014xxa} and for defect CFT in \cite{Kobayashi:2018lil}, it was proposed in \cite{Giombi:2020rmc} that the rescaled free energy on AdS with a sphere boundary
\begin{equation} \label{FTildeDefn}
\tilde{F} = - \sin \left( \frac{\pi (d-1)}{2} \right) F_{AdS_d} 
\end{equation}
should decrease under RG flows localized on the boundary: $\tilde{F}_{\rm UV}>\tilde F_{\rm IR}$.\footnote{Here we are only making a statement about the difference of $\tilde F$ between the UV and IR boundary fixed points, and not about the value of $\tilde F$ along the flow.} We check in section \ref{Sec:LargeN} by computing the AdS free energy for the various boundary conformal phases that it indeed does satisfy such inequality under the boundary RG flow.  

In the case of $d=3$, a very interesting extension that we leave to future work would be to gauge the $U(N)$ global symmetry and couple the fermions to the Chern-Simons gauge theory. One may consider adding Chern-Simons interactions either in the model of $N$ free fermions, or at the critical point of the Gross-Neveu model. Similarly, one may consider gauging  
the scalar CFTs with a Chern-Simons term (either in the critical model, or starting with the free scalar theory without quartic interaction). Then, one may study how the bose-fermi dualities \cite{Giombi:2011kc, Aharony:2011jz, Aharony:2012nh, Aharony:2015mjs, Hsin:2016blu} (see also \cite{Giombi:2016ejx} for a review) are realized in the presence of a boundary. An interesting observation, which was also pointed out in \cite{Carmi:2018qzm}, is that in $d =3$, the large $N$ dimensions of the leading boundary fermion in the two phases $B_1$ and $B_2$ of the GN model are $3/2$ and $1/2$ respectively, see Table  \ref{TableBoundaryPhasesDimensions}. These coincide with the dimensions of the leading boundary scalar in a free massless boson theory with Dirichlet or Neumann boundary conditions, respectively. On the other hand, for the free massless fermion, there is just one phase, with leading boundary fermion of dimension $1$, which happens to be the same as the dimension of the leading boundary fundamental scalar in the so-called ordinary transition in the large $N$ scalar BCFT (see \cite{Giombi:2020rmc}). The fact that the dimensions of the boundary fundamental fermionic and bosonic operators match this way should be related to the bose-fermi duality, and suggests that the Chern-Simons interactions may not affect those boundary scaling dimensions to leading order at large $N$. It would be interesting to clarify this, as well as compute other observables in the Chern-Simons scalar and fermion theories, like the AdS$_3$ free energy (which encodes the boundary conformal anomaly), and boundary four-point functions of the fundamental fields.  

The rest of this paper is organized as follows: We start in section \ref{Sec:FreeMassiveFermion} by studying a single free massive fermion in AdS. We calculate the bulk and boundary two-point function of the fermion, and study possible boundary conditions and the AdS free energy for these boundary conditions. Then in section \ref{Sec:LargeN}, we study large $N$ $U(N)$ Gross-Neveu model and discuss the various phases we described above. In section \ref{Sec:EpsExp}, we describe these phases in the GN model in $d = 2 + \epsilon$ and in the GNY model in $d = 4 - \epsilon$. Finally, in section \ref{Sec:EOM}, we calculate the bulk two-point functions to leading order in $\epsilon$ in both GN and GNY models and compare the results with those of the large $N$ expansion. In particular, following an approach proposed in \cite{Giombi:2020rmc}, we use bulk equations of motion to derive a differential equation that the two-point function must satisfy, and then solve it to extract the bulk two-point function. 

\section{Free massive fermion on hyperbolic space} \label{Sec:FreeMassiveFermion}
Let's start by reviewing some facts about free fermions on hyperbolic space to set the notation. This is mostly a review and the material is well discussed \cite{Henneaux:1998ch, Mueck:1998iz, Arutyunov:1998ve, Mueck:1999efk, Henningson:1998cd, Kawano:1999au, Basu:2006ti, Aharony:2010ay, Aros:2011iz, Carmi:2018qzm, Sato:2021eqo}. We start with the following Euclidean action
\begin{equation}
S = - \int d^d x \sqrt{g} \bar{\Psi} (\gamma \cdot \nabla + m ) \Psi .
\end{equation}
As discussed in \cite{Henneaux:1998ch, Allais:2010qq, Laia:2011wf}, one needs to add a boundary term to this action to have a well defined variational principle. But we will not need it so we do not write it down explicitly. For the most part, we will use Poincar\'e coordinates $ (x^0, x^i) = (z, \textbf{x})$, $i = 1,..., d-1$ with the metric 
\begin{equation}
ds^2 = \frac{dz^2 + d \textbf{x}^2}{z^2}.
\end{equation}
In these coordinates, the vielbein $e^{\mu}_a = z \delta^{\mu}_a$, so that $ \gamma \cdot \nabla = z \gamma^a \nabla_a $ with $\gamma^a$ being the flat space gamma matrices. The spin connection and the Dirac operator take the following form 
\begin{equation}
\omega^{ab}_{\mu} = \frac{\delta^a_0 \delta^b_{\mu} - \delta^b_0 \delta^a_{\mu}}{z}, \hspace{0.5cm} \gamma \cdot \nabla \Psi = e^{\mu}_a \gamma^a \left( \partial_{\mu} + \frac{\omega^{bc}_{\mu} [\gamma_b, \gamma_c]}{8} \right) \Psi = \left( z \gamma^a \partial_a - \frac{(d - 1)}{2} \gamma_0   \right) \Psi .
\end{equation} 
Also note that we are using $0$ for the radial ($z$) direction, so $\bar{\Psi} = \Psi^{\dagger} \gamma^i$ with $i$ being set equal to the Euclidean time direction. 

At the boundary of the hyperbolic space, one can impose on the fermion two possible boundary conditions $\gamma_0 \Psi(z \rightarrow 0, \textbf{x}) = \pm \Psi(z \rightarrow 0, \textbf{x})$ or equivalently ($\bar{\Psi} \gamma_0  = \mp \bar{\Psi}$) which we will refer to as $+$ and $-$ boundary condition respectively. Let us first review the calculation of the fermion two-point function $\langle \Psi (x_1) \bar{\Psi} (x_2) \rangle = G_{\Psi} (x_1, x_2)$, which can be found by solving the following differential equation 
\begin{equation} 
(\gamma \cdot \nabla + m) G_{\Psi} (x_1, x_2) = - \delta^d (x_1 - x_2).
\end{equation}
To solve it, we start with the following ansatz \cite{Mueck:1999efk, Basu:2006ti}
\begin{equation} \label{FermionTwoPointAnsatz}
G_{\Psi} (x_1, x_2) = \left(  - \frac{\gamma_0 \gamma_{a} (\bar{x}_1 - x_2)^{a} }{ \sqrt{z_1 z_2}} \frac{\alpha(\zeta)}{\sqrt{\zeta + 4}} + \frac{\gamma_{a} (x_1 - x_2)^{a} }{\sqrt{z_1 z_2}} \frac{\beta(\zeta)}{\sqrt{\zeta}} \right)
\end{equation}
where the cross-ratio $\zeta$ is defined by 
\begin{equation}
\zeta = \frac{\textbf{x}_{12}^2 + z_{12}^2}{z_1 z_2}
\end{equation}
and $\bar{x} = (-z, \mathbf{x})$ is the image point with respect to the boundary. We then act on the ansatz with the Dirac operator which gives
\begin{equation} \label{DiracOperatorAnsatz}
\begin{split}
&\gamma \cdot \nabla_1  G_{\Psi} (x_1, x_2) = \left( z_1 \gamma^a \partial_{1a} - \frac{(d - 1)}{2} \gamma_0 \right) G_{\Psi} (x_1, x_2) \\
&= \frac{\gamma_{a} x_{12}^{a} \sqrt{\zeta + 4} }{\sqrt{z_1 z_2}} \left( \alpha'(\zeta) + \frac{(d - 1)}{2} \frac{\alpha (\zeta)}{\zeta + 4} \right)  - \frac{\gamma_0 \gamma_{a} \bar{x}_{12}^{a} \sqrt{\zeta} }{ \sqrt{z_1 z_2}} \left( \beta'(\zeta) + \frac{(d - 1)}{2} \frac{\beta (\zeta)}{\zeta} \right).
\end{split}
\end{equation}
Hence, the massive Dirac equation on this ansatz gives following set of coupled equations 
\begin{equation}
\begin{split}
&\alpha'(\zeta) + \frac{d-1}{2} \frac{\alpha (\zeta)}{4 + \zeta} = - \frac{m \beta(\zeta)}{\sqrt{\zeta (4 + \zeta)}} \\
& \beta'(\zeta) +  \frac{d-1}{2} \frac{\beta (\zeta)}{\zeta} = - \frac{ m \alpha(\zeta)}{\sqrt{\zeta (4 + \zeta)}}.
\end{split}
\end{equation}
We can solve it by substituting for $\beta(\zeta)$ from the first equation into the second one, which gives a second order equation for $\alpha(\zeta)$. This has two solutions, which gives two choices of propagator corresponding to two possible boundary fall-offs. The first one has a leading boundary fermion of dimension $(d-1)/2 + |m|$ in its boundary spectrum 
\begin{equation} \label{FermionMassiveTwoPMin}
\begin{split}
G_{\Psi} (x_1, x_2) = & \frac{-\Gamma\left( \frac{d}{2} + |m| \right)}{\Gamma \left( \frac{1}{2} + |m| \right) 2 \pi^{\frac{d-1}{2}}} \bigg[ \frac{\gamma_{a} x_{12}^{a} }{\sqrt{z_1 z_2} } \frac{(4 + \zeta)^{1 - \frac{d}{2}}}{\zeta^{|m| + 1} } {}_2 F_1 \left( 1 + |m| - \frac{d}{2} , 1 + |m|, 1 + 2 |m|, - \frac{4}{\zeta} \right)  \\
&- \textrm{sgn}(m) \frac{\gamma_0 \gamma_{a} (\bar{x}_{12})^{a} }{ \sqrt{z_1 z_2}} \frac{1}{\zeta^{|m|} (4 + \zeta)^{\frac{d}{2}}} {}_2 F_1 \left( 1 + |m| - \frac{d}{2} , |m|, 1 + 2 |m|, - \frac{4}{\zeta} \right)    \bigg].
\end{split}
\end{equation}
This is allowed for all values of mass and is known in the literature as standard quantization. This satisfies the boundary condition $\gamma_0 G_{\Psi} (x_1, x_2)|_{z_1 \rightarrow 0} = - \textrm{sgn}(m) G_{\Psi} (x_1, x_2)|_{z_1 \rightarrow 0} $. We can set $z_1$ or $z_2 = 0$ to get the bulk-boundary two-point function of the fermion. Defining the boundary spinors of dimension $\hat{\Delta} = \frac{d-1}{2} + |m|$ as $\hat{\Psi} (\textbf{x}) = z^{- \hat{\Delta}} \Psi (\textbf{x}, z \rightarrow 0)$, we have 
\begin{equation}\label{FermionBulkBoundaryGen}
\begin{split}
\langle \hat{\Psi} (\textbf{x}_1) \bar{\Psi} (x_2) \rangle &= -  \left( \frac{1 - \textrm{sgn}(m) \gamma_0}{2} \right) \frac{(\gamma_a x_{12}^a) \Gamma \left(\hat{\Delta} + \frac{1}{2} \right) }{\sqrt{z_2} \pi^{\frac{d - 1}{2}}  \Gamma \left(\hat{\Delta} - \frac{d}{2} + 1 \right) } \left( \frac{z_2}{z_2^2 + \textbf{x}_{12}^2} \right)^{\hat{\Delta} + \frac{1}{2} } \\
\langle \Psi (x_1) \hat{\bar{\Psi}} (\textbf{x}_2) \rangle &= - \frac{(\gamma_a x_{12}^a) \Gamma \left(\hat{\Delta} + \frac{1}{2} \right) }{\sqrt{z_1} \pi^{\frac{d - 1}{2}}  \Gamma \left(\hat{\Delta} - \frac{d}{2} + 1 \right) } \left( \frac{z_1}{z_1^2 + \textbf{x}_{12}^2} \right)^{\hat{\Delta} + \frac{1}{2} } \left( \frac{1 + \textrm{sgn}(m) \gamma_0}{2} \right).
\end{split}
\end{equation} 

The other possible boundary fall-off is when the leading boundary spinor has dimension $(d-1)/2 - |m|$. This is only unitary for $|m| < 1/2$ and is known in the literature as alternative quantization. The corresponding two-point function is 

\begin{equation} \label{FermionMassiveTwoPPlus}
\begin{split}
G_{\Psi} (x_1, x_2) = & \frac{-\Gamma\left( \frac{d}{2} - |m| \right)}{\Gamma \left( \frac{1}{2} - |m| \right) 2 \pi^{\frac{d-1}{2}}} \bigg[ \frac{\gamma_{a} x_{12}^{a} }{\sqrt{z_1 z_2} } \frac{(4 + \zeta)^{1 - \frac{d}{2}}}{\zeta^{-|m| + 1} } {}_2 F_1 \left( 1 - |m| - \frac{d}{2} , 1 - |m|, 1 - 2 |m|, - \frac{4}{\zeta} \right)  \\
& + \textrm{sgn}(m) \frac{\gamma_0 \gamma_{a} (\bar{x}_{12})^{a} }{ \sqrt{z_1 z_2}} \frac{1}{\zeta^{-|m|} (4 + \zeta)^{\frac{d}{2}}} {}_2 F_1 \left( 1 - |m| - \frac{d}{2} , -|m|, 1 - 2 |m|, - \frac{4}{\zeta} \right)    \bigg].
\end{split}
\end{equation}
This satisfies the boundary condition $\gamma_0 G_{\Psi} (x_1, x_2)|_{z_1 \rightarrow 0} = \textrm{sgn}(m)  G_{\Psi} (x_1, x_2)|_{z_1 \rightarrow 0} $. In the massless limit, $m = 0$, the two cases become degenerate with the propagator given by \footnote{This is related, by a Weyl transformation, to the result in flat space written in \cite{McAvity:1993ue}, if we pick $U = \pm \gamma_0$ and $\bar{U} = \mp \gamma_0$.}
\begin{equation} \label{MasslessProp1}
G_{\Psi} (x_1, x_2) = - \frac{\Gamma\left( \frac{d}{2} \right)}{2 \pi^{\frac{d}{2}}} \bigg[ \frac{\gamma_{a} x_{12}^{a} }{\sqrt{z_1 z_2} } \frac{1}{\zeta^{\frac{d}{2}}} \pm \frac{\gamma_0 \gamma_{a} (\bar{x}_{12})^{a} }{ \sqrt{z_1 z_2}} \frac{1}{(4 + \zeta)^{\frac{d}{2}}} \bigg]
\end{equation}
which satisfies the boundary condition $\gamma_0 G_{\Psi} (x_1, x_2)|_{z_1 \rightarrow 0} = \pm  G_{\Psi} (x_1, x_2)|_{z_1 \rightarrow 0} $.

\subsection{Boundary correlation functions } \label{SubSec:BoundaryProjection}
In this subsection, we explain how to obtain correlation functions in the boundary theory from the bulk. As a first step, we need to take the boundary limit of \eqref{FermionBulkBoundaryGen} 
\begin{equation} \label{BulkRepTwoP} 
\langle \hat{\Psi} (\textbf{x}_{1})  \hat{\bar{\Psi}}(\textbf{x}_{2}) \rangle =  -\frac{\Gamma \left(\hat{\Delta} + \frac{1}{2} \right)}{2 \pi^{\frac{d-1}{2}} \Gamma \left(\hat{\Delta} - \frac{d}{2} + 1 \right) } \frac{ (1 \pm \gamma_0) \boldsymbol{\gamma} \cdot \textbf{x}_{12}}{(\textbf{x}_{12}^2)^{ \hat{\Delta} + \frac{1}{2}}}.
\end{equation}
Note that the fermion operators on the boundary have half as many components as the ones in the bulk, because the boundary condition sets the other half to $0$. So we need to project the above two-point function onto the boundary fermion representation. When the bulk is even, the boundary fermions are Dirac fermions, while when the bulk is odd, the boundary fermions are Weyl.  
Let us start with the case when the bulk is even dimensional. For concreteness, let us choose the following representation of Dirac matrices 
\begin{equation}
\gamma_0 = \begin{pmatrix}
\mathbb{I} & 0 \\
0 & - \mathbb{I}
\end{pmatrix}, \hspace{1cm} \gamma_i =  \begin{pmatrix}
0 & \Gamma_i \\
\Gamma_i & 0
\end{pmatrix}
\end{equation}
where $\Gamma_i$ are the Dirac matrices in $d - 1$ dimensions and $\mathbb{I}$ is the $c_{d-1} \times c_{d-1}$ dimensional identity. We defined $c_d = 2^{\lfloor \frac{d}{2} \rfloor}$ as the number of components of a Dirac spinor in $d$ dimensions. We now only restrict to $+$ boundary condition on the fermion and $m > 0$ (the other cases are identical), in which case, we can choose the following bulk polarization spinors
\begin{equation} \label{BoundaryPolarization}
S = \begin{pmatrix}
0 \\
v
\end{pmatrix}, \hspace{1cm} \bar{S} =  \begin{pmatrix}
\bar{v} & 0
\end{pmatrix}
\end{equation}
where $v$ is the boundary polarization spinor. We can then define the boundary fermion operator by $\bar{v} \psi (\textbf{x}) = \bar{S} \hat{\Psi} (\textbf{x})$. Contracting the two-point function with these polarization spinors, we get
\begin{equation}
\langle \bar{v}_1 \psi (\textbf{x}_1) \bar{\psi} (\textbf{x}_2) v_2  \rangle =  \langle \bar{S}_1 \hat{\Psi} (\textbf{x}_{1})  \hat{\bar{\Psi}}(\textbf{x}_{2}) S_2 \rangle =  -\frac{\Gamma \left(\hat{\Delta} + \frac{1}{2} \right)}{ \pi^{\frac{d-1}{2}} \Gamma \left(\hat{\Delta} - \frac{d}{2} + 1 \right) } \frac{  \bar{v}_1 \boldsymbol{\Gamma} \cdot \textbf{x}_{12} v_2 }{(\textbf{x}_{12}^2)^{ \hat{\Delta} + \frac{1}{2}}}.
\end{equation}
We can then differentiate with respect to boundary polarization spinors to get the correlation functions on the boundary 
\begin{equation} \label{BoundaryRepTwoP}
\langle \psi (\textbf{x}_1) \bar{\psi} (\textbf{x}_2) \rangle =  -\frac{\Gamma \left(\hat{\Delta} + \frac{1}{2} \right)}{ \pi^{\frac{d-1}{2}} \Gamma \left(\hat{\Delta} - \frac{d}{2} + 1 \right) } \frac{ \boldsymbol{\Gamma} \cdot \textbf{x}_{12} }{(\textbf{x}_{12}^2)^{ \hat{\Delta} + \frac{1}{2}}}.
\end{equation}
In the free theory, the higher point functions can then be just constructed by Wick contractions. However, when the bulk has additional interactions, as we will show in section \ref{Sec:EpsExp}, we should start with fermions in the bulk representation, and then project onto the boundary fermion representation. 

We now comment on what happens when the bulk is odd dimensional. In this case, the Dirac matrices on the boundary have the same dimension as the bulk and are just given by the bulk gamma matrices $\gamma_i$ with $\gamma_0$ being the chirality matrix. The boundary fermion operator is a Weyl fermion and we can take it to be just $\hat{\Psi} (\textbf{x})$ satisfying $\gamma_0 \hat{\Psi} (\textbf{x}) = \pm \hat{\Psi} (\textbf{x})$. The two-point function is given by \eqref{BulkRepTwoP}. An immediate consequence of the fact that the fermion is Weyl is that for a single Dirac fermion in the bulk, the leading boundary scalar $ \hat{\bar{\Psi}} \hat{\Psi} (\textbf{x})$ vanishes. So the leading boundary scalar should have dimension $d - 2 m$  instead of $d-1 - 2m$ and should include a derivative.  

\subsection{Free energy}
Next we calculate the free energy on hyperbolic space. To do that, we compactify the boundary of the hyperbolic space to a sphere. The free energy is then given by the following trace
\begin{equation}
F = - \textrm{tr} \log \left( \gamma \cdot \nabla + m \right). 
\end{equation} 
We need to know the spectrum of Dirac operator on hyperbolic space \cite{Camporesi:1995fb, Bytsenko:1994bc}. The eigenvalues of $\gamma \cdot \nabla$ are $\pm i \lambda $ with the degeneracy given by 
\begin{equation}
\mu(\lambda) = \frac{\textrm{Vol} (H^d) c_d}{(4 \pi)^{\frac{d}{2}} \Gamma \left( \frac{d}{2} \right) } \bigg| \frac{\Gamma \left( \frac{d}{2} + i \lambda \right)}{\Gamma \left( \frac{1}{2} + i \lambda \right) } \bigg|^2.
\end{equation}
The free energy does not depend on the sign of $m$, so we will just take $m > 0$ for this calculation. Using the above results, the free energy is given by the following spectral integral
\begin{equation} \label{FreeEnergyMassive}
\begin{split}
F &= -  \frac{ \textrm{Vol} (H^d) c_d}{(4 \pi)^{\frac{d}{2}} \Gamma \left( \frac{d}{2} \right) } \int_0^{\infty} d \lambda \frac{ | \Gamma \left( \frac{d}{2} + i \lambda \right)|^2 \log \left( \lambda^2 + m^2 \right) }{|\Gamma \left( \frac{1}{2} + i \lambda \right)|^2 } \\
&= \frac{ \textrm{Vol} (H^d) c_d }{(4 \pi)^{\frac{d}{2}} \Gamma \left( \frac{d}{2} \right)} \frac{\partial}{\partial \alpha} \left[ \int_0^{\infty} d \lambda \frac{ | \Gamma \left( \frac{d}{2} + i \lambda \right)|^2 }{|\Gamma \left( \frac{1}{2} + i \lambda \right)|^2 \left( \lambda^2 + m^2 \right)^{\alpha}}  \right] \bigg|_{\alpha \rightarrow 0}
\end{split}
\end{equation}
The above integral is hard to do analytically for arbitrary $d$, but can be performed if we plug in $d = 3$
\begin{equation} \label{FreeEnergy3dGen}
F (\hat{\Delta}) = \frac{ \textrm{Vol} (H^3) (\hat{\Delta} - 1) \left( 4 (\hat{\Delta} - 1)^2 - 3  \right) }{24 \pi}.
\end{equation} 
where we wrote the answer in terms of $\hat{\Delta} = (d-1)/2 + m$. Even though we used the $+$ sign, the final result can be analytically continued for both $\hat{\Delta} = (d-1)/2 \mp m$. For the free massless fermion, $\hat{\Delta} = 1$, so $F = 0$. For $d \neq 3$, the integral can be performed if we first take a derivative with $m$
\begin{equation}
\begin{split}
\frac{\partial F}{\partial m} &= - \textrm{tr} \left( \frac{1}{\gamma \cdot \nabla + m } \right) =  -  \frac{ \textrm{Vol} (H^d)  m  c_d}{(4 \pi)^{\frac{d}{2}} \Gamma \left( \frac{d}{2} \right) } \int_{-\infty}^{\infty} d \lambda \frac{ | \Gamma \left( \frac{d}{2} + i \lambda \right)|^2 }{|\Gamma \left( \frac{1}{2} + i \lambda \right)|^2 \left( \lambda^2 + m^2 \right) } \\
&= -  \frac{ \textrm{Vol} (H^d) c_d \Gamma \left( 1 - \frac{d}{2} \right) \Gamma \left( \frac{d}{2} + m \right)  }{(4 \pi)^{\frac{d}{2}} \Gamma \left(1 - \frac{d}{2} + m \right) }. 
\end{split}
\end{equation}
We did the above integral by closing the contour in the upper half $\lambda$ plane and summing over the residues at $\lambda = i m$ and at $\lambda = i (d/2 + n)$ for $n \geq 0$. The arc at infinity can only be dropped for $d < 2$, but the final result can be analytically continued to $d > 2$. This trace can also be obtained by taking the short distance limit and tracing over the two-point function in \eqref{FermionMassiveTwoPMin}.

As a side remark, we note that for the mass range $0 < |m| < 1/2$ where both the boundary conditions are allowed, they are related by a RG flow on the boundary triggered by a fermionic bilinear \cite{Amsel:2008iz, Allais:2010qq, Laia:2011wf, Aros:2011iz}. The fermion bilinear in alternate quantization is relevant with scaling dimension $d - 1 - 2 m$ and may be used to flow to the standard quantization \cite{Laia:2011wf} \footnote{This flow is not possible for a single bulk fermion in odd dimensions. Because in that case, the boundary fermion is Weyl, and hence the leading bilinear scalar vanishes.}. There is a general formula for the free energy change under a flow by the square of a spin $1/2$ single-trace operator in a CFT that obeys large $N$ factorization \cite{Giombi:2014xxa, Allais:2010qq, Aros:2011iz} \footnote{Note that when the bulk is odd dimensional, our formula differs from that of \cite{Giombi:2014xxa} by a factor of $2$. This is because the result in \cite{Giombi:2014xxa} is given for a Dirac fermion, whereas in our case, when bulk is odd, the boundary condition forces the boundary spinor to be a Weyl fermion, which has half the number of components as that of a Dirac fermion.}
\begin{equation}
F_{d - 1- \hat{\Delta}} - F_{\hat{\Delta}} = - \frac{ c_d}{\sin \left( \frac{\pi (d-1)}{2}\right) \Gamma(d)} \int_0^{\hat{\Delta} - \frac{d-1}{2}} d u \cos (\pi u) \Gamma \left( \frac{d}{2} + u \right) \Gamma \left( \frac{d}{2} - u \right). 
\end{equation}
In the AdS/CFT context, this corresponds to the difference in free energy between the same bulk theory with the two possible boundary conditions for the bulk fermion dual to the boundary single-trace operator. Even though in our case the flow between the two boundary conditions in the free fermion theory is not a double-trace flow in the usual sense, mathematically the problem is equivalent and we can still calculate the free energy difference between the two boundary conditions using the above formula. In $d = 3$, it gives 
\begin{equation}
F_{2- \hat{\Delta}} - F_{\hat{\Delta}} = -\frac{\textrm{Vol} (H^3) (\hat{\Delta} - 1) (1 + 4 \hat{\Delta} (\hat{\Delta} - 2)) }{12 \pi}
\end{equation}
where we used the fact that the regularized volume of hyperbolic space is given by $\textrm{Vol} (H^d) = \pi^{\frac{d - 1}{2}} \Gamma\left( \frac{1 - d}{2} \right)$. This agrees with what we get by using the explicit result for $d = 3$ free energy \eqref{FreeEnergy3dGen}.

As was discussed in \cite{Giombi:2020rmc}, the free energy on hyperbolic space is also related to the trace anomaly coefficients. In $d = 3$, on manifolds with a boundary, the trace anomaly is given by \cite{Graham:1999pm, Jensen:2015swa, Herzog:2017kkj}
\begin{equation} 
\langle {T^{\mu}}_{\mu} \rangle^{d = 3} = \frac{\delta(x_{\perp})}{4 \pi} \left( a_{3 d} \hat{{\cal R}} + b \tr \hat{K}^2 \right).
\end{equation}
In the above equation, $\hat{{\cal R}}$ is the boundary Ricci scalar and $\hat{K}_{ij}$ is the traceless part of the extrinsic curvature associated to the boundary. Following the logic in \cite{Giombi:2020rmc}, it can be shown that for free massive fermions, the coefficient $a_{3d}$ is given by 
\begin{equation} \label{a3dDef}
a_{3d} = \frac{ (\hat{\Delta} - 1) \left( 4 (\hat{\Delta} - 1)^2 - 3  \right) }{24 }.
\end{equation}
This vanishes for massless fermions, in agreement with the results in \cite{Fursaev:2016inw}. In what follows, we will also calculate this anomaly coefficient for large $N$ interacting fixed points. It should also be possible to extract this coefficient from the fermion free energy on a round ball, which was calculated for free fermions in \cite{Dowker:1995sw, Rodriguez-Gomez:2017aca}.

\section{Large $N$ Gross-Neveu model} \label{Sec:LargeN}
In this section, we study the Gross-Neveu model for $N$ interacting Dirac fermions in AdS \cite{Carmi:2018qzm} and do perturbation theory in $1/N$. Starting with the action \eqref{ActionGNLarge}, we can integrate out the fermions to get an effective action in terms of $\sigma$
\begin{equation} \label{EffectiveAction}
Z = e^{-F} = \int [d \sigma] [d \Psi^i] [d \bar{\Psi}_i] e^{\int d^d x \sqrt{g_x} \left( \bar{\Psi}_i \gamma \cdot \nabla \Psi^i + \sigma \bar{\Psi}_i \Psi^i \right)}   = \int [d \sigma] \exp \left( N \mathrm{tr} \log (\gamma \cdot \nabla + \sigma(x)) \right).
\end{equation}
At leading order in large $N$, the path integral over $\sigma$ may be performed by a saddle point approximation, assuming a constant saddle at $\sigma(x) = \sigma^*$. This constant can be found by solving 
\begin{equation} 
\frac{\partial F}{\partial \sigma^*} = - N \textrm{tr} \left[ \frac{1}{\gamma \cdot \nabla + \sigma^*} \right] = 0.
\end{equation} 
It is clear that at this order, $\sigma^*$ acts like a mass for the fermions. The trace above can be obtained from the two-point function \eqref{FermionMassiveTwoPMin} or \eqref{FermionMassiveTwoPPlus} depending upon the boundary fall-off for the fermion. For the boundary condition $\gamma_0 G_{\Psi} (x_1, x_2)|_{z_1 \rightarrow 0} = - \textrm{sgn}(\sigma^*) G_{\Psi} (x_1, x_2)|_{z_1 \rightarrow 0} $, using \eqref{FermionMassiveTwoPMin}
\begin{equation} \label{FreeEneryDer2}
\frac{\partial F}{\partial \sigma^*} = - N \textrm{tr} \left[ \frac{1}{\gamma \cdot \nabla + \sigma^*} \right] = -\frac{N \textrm{sgn} (\sigma^*)  c_d \textrm{Vol} (H^d)}{(4 \pi)^{\frac{d}{2}}} \frac{\Gamma \left( 1- \frac{d}{2} \right) \Gamma \left( \frac{d}{2} + |\sigma^*| \right) }{ \Gamma \left( 1 - \frac{d}{2} + |\sigma^*| \right)}.
\end{equation} 
This gives the following large $N$ saddle 
\begin{equation}
|\sigma^*| = \frac{d}{2} - 1 - n \implies \hat{\Delta} = d - \frac{3}{2} - n
\end{equation}
for a non-negative integer $n$. For $2 < d < 4$, there is only one possible solution
\begin{equation} \label{LargeNDimPhase1}
|\sigma^*| = \frac{d}{2} - 1 \implies \hat{\Delta} = d - \frac{3}{2}.
\end{equation}
The unitarity bound at the boundary requires $\hat{\Delta} \geq (d-2)/2$, which is satisfied for all $2 < d < 4$. This is the phase we called $B_1$ in the introduction. 

For the other boundary condition $\gamma_0 G_{\Psi} (x_1, x_2)|_{z_1 \rightarrow 0} =  \textrm{sgn}(\sigma^*) G_{\Psi} (x_1, x_2)|_{z_1 \rightarrow 0} $, we have, using \eqref{FermionMassiveTwoPPlus} 
\begin{equation} \label{FreeEneryDer1}
\frac{\partial F}{\partial \sigma^*} = - N  \textrm{tr} \left[ \frac{1}{\gamma \cdot \nabla + \sigma^*} \right] = \frac{N \textrm{sgn} (\sigma^*)  c_d \textrm{Vol} (H^d)}{(4 \pi)^{\frac{d}{2}}} \frac{\Gamma \left( 1- \frac{d}{2} \right) \Gamma \left( \frac{d}{2} - |\sigma^*| \right) }{ \Gamma \left( 1 - \frac{d}{2} - |\sigma^*| \right)}.
\end{equation} 
This gives the following large $N$ saddle 
\begin{equation}
|\sigma|^* = -\frac{d}{2} + 1 + n \implies \hat{\Delta} = d - \frac{3}{2} - n
\end{equation}
for a positive integer $n$. There is no unitary saddle for $d < 3$, while in $3 \leq d \leq 4$, $n = 1$ gives a unitary saddle
\begin{equation} \label{LargeNDimPhase2} 
|\sigma^*| = 2 - \frac{d}{2}   \implies \hat{\Delta} = d - \frac{5}{2} .
\end{equation}
This is the saddle for both $B_2$ and $B_2'$ phases, and as we show below, the two phases can only be distinguished by the $\sigma$ fluctuations around this saddle which are subleading in $1/N$. 

Let us also write explicitly, the fermion two-point function for the two cases. Plugging in $|\sigma^*| = \frac{d}{2} - 1$ into \eqref{FermionMassiveTwoPMin}, we get
\begin{equation} \label{LargeNFermTwoPP1}
G^{|\sigma^*| = \frac{d}{2} - 1}_{\Psi} (x_1, x_2) = -\frac{2^{d-3} \Gamma \left(\frac{d}{2} \right)}{ \pi^{\frac{d}{2}} \left( \zeta (4 + \zeta)\right)^{\frac{d}{2}} \sqrt{z_1 z_2}} \left( \gamma \cdot (x_1 - x_2) (4 + \zeta) - \textrm{sgn} (\sigma^*) \gamma_0 \ \gamma \cdot (\bar{x}_1 - x_2) \zeta \right). \\
\end{equation}
As we show below, in $d = 2 + \epsilon$, this saddle should match with the calculation in an $\epsilon$ expansion in Gross-Neveu model. The negative value of $\sigma^*$, i.e. $\sigma^* = 1 - \frac{d}{2} $ matches the $\epsilon$ expansion calculation if we do perturbation theory around free theory with a $+$ boundary condition on the fermion, and similarly for the other sign. This is consistent with the boundary condition obeyed by the propagator we write here i.e. a negative $\sigma^*$ gives a $+$ boundary condition for the fermion and vice versa. In $d = 4 - \epsilon$, this saddle matches to a phase in Gross-Neveu-Yukawa model where the scalar gets a classical vev.

For the other saddle, we plug in $|\sigma^*| =  2 - \frac{d}{2}$ into \eqref{FermionMassiveTwoPPlus}, and we get
\begin{equation} \label{LargeNFermTwoPP2}
\begin{split}
G^{|\sigma^*| =  2 - \frac{d}{2}}_{\Psi} (x_1, x_2) = - \frac{2^{d-5} \Gamma \left(\frac{d}{2} - 1 \right)}{  \pi^{\frac{d}{2}} \left( \zeta (4 + \zeta)\right)^{\frac{d}{2}} \sqrt{z_1 z_2}} \bigg[& \gamma \cdot (x_1 - x_2) (4 + \zeta) \left( d (2 + \zeta) - 3 \zeta - 4\right) + \\
& \textrm{sgn} (\sigma^*) \gamma_0 \ \gamma \cdot (\bar{x}_1 - x_2) \zeta \left( d (2 + \zeta) - 3 \zeta - 8 \right) \bigg].
\end{split}
\end{equation}
In $d = 4 -\epsilon$, this saddle matches with the $\epsilon$ expansion calculation in GNY model where the scalar does not get a classical vev. The positive value of $\sigma^* = 2 - \frac{d}{2}$ matches the perturbation theory around the free theory with $+$ boundary condition. This again, is consistent with the propagator we write here. Note that the two signs of $\sigma^*$ give two essentially equivalent theories. They only differ by the signs of one-point functions of parity-odd operators.

\subsection{$\sigma$ fluctuations}
In this subsection, we consider fluctuations about the constant $\sigma$ saddles that we found above. So we expand the effective action in \eqref{EffectiveAction} about the constant $\sigma$ background $\sigma(x) = \sigma^* + \delta \sigma(x)$
\begin{equation}
\begin{split}
S_{\textrm{eff}} (\sigma) &= - N \mathrm{tr} \log \left( \gamma \cdot \nabla + \sigma^* + \delta \sigma(x) \right) = - N \mathrm{tr} \log \left( \gamma \cdot \nabla + \sigma^* \right) + \frac{N}{2} \mathrm{tr} \left( \frac{ \delta \sigma}{ \gamma \cdot \nabla + \sigma^* } \right)^2\\
&= - N \mathrm{tr} \log \left( \gamma \cdot \nabla + \sigma^* \right) + \frac{N}{2} \int d^d x d^d y \sqrt{g_x} \sqrt{g_y} \ \mathrm{Tr} \left[ G_{\Psi} (x,y) G_{\Psi} (y,x)  \right] \delta \sigma(x) \delta \sigma(y)
\end{split}
\end{equation}
where $\mathrm{Tr}$ is the trace over the fermionic indices while $\mathrm{tr}$ includes trace over both spacetime and fermionic indices. The $\sigma$ propagator can then be read off from the inverse of the quadratic piece 
\begin{equation} \label{SigmaPropInversion}
\int d^d x_3 \sqrt{g} \ \mathrm{Tr} \left[ G_{\Psi}(x_1, x_3) G_{\Psi}(x_3, x_1) \right] G_{\sigma} (x_3, x_2) =  \frac{1}{N} \frac{\delta^d (x_1 - x_2) }{\sqrt{g_{x_1}}}.
\end{equation}
For $\sigma^* = d/2 - 1$, we need to invert 
\begin{equation} 
\mathrm{Tr} \left[ G_{\Psi}(x_1, x_3) G_{\Psi}(x_3, x_1) \right] = -\frac{ 4^{d} \Gamma \left(\frac{d}{2} \right)^2 c_d}{16 \pi^{d} \left( \zeta (4 + \zeta)\right)^{d-1}}
\end{equation}
while for $\sigma^* = d/2 - 2$, we need to find the inverse of 
\begin{equation}
\mathrm{Tr} \left[ G_{\Psi}(x_1, x_3) G_{\Psi}(x_3, x_1) \right] = -\frac{ 4^{d}  \Gamma \left(\frac{d}{2} - 1 \right)^2 c_d}{64 \pi^{d} \left( \zeta (4 + \zeta)\right)^{d-1}} \left( (d-2)^2 +  \frac{(d-1)(d-3)}{4} \zeta (4 + \zeta) \right).
\end{equation}
We give the details of this inversion in the appendix \ref{AppendixSigmaProp} and just report the result here. For $\sigma^* = d/2 - 1$ i.e. $B_1$ phase, we find \eqref{SigmaCorrAppP1} \footnote{Note that this is only the connected piece of the $\sigma$ two-point function, so that the complete two-point function is $(\sigma^*)^2 + G_{\sigma}$}
\begin{equation} \label{SigmaCorrMainB1}
G_{\sigma} (\zeta) =  -\frac{2^{4 d-5} (d-2) \Gamma \left(\frac{d-1}{2}\right)^2 \Gamma (d)}{ N c_d \pi  \Gamma \left(\frac{d}{2}\right) \Gamma \left(1-\frac{d}{2}\right) \Gamma (2 d-2) \zeta^d} \ {}_2 F_1 \left(d, d- 1, 2 d - 2 , -\frac{4}{\zeta} \right).
\end{equation}

The two-point function of a scalar operator $O$ in a BCFT can be expanded into bulk and boundary channel conformal blocks as \cite{McAvity:1995zd, Billo:2016cpy}
\begin{equation} \label{BlockDecompTwoPS}
G_{O} (\zeta) = \frac{A}{\zeta^{\Delta_{O}}} \left( 1 + \sum_k \lambda_k f_{\textrm{bulk}} \left( \Delta_k; \zeta \right)  \right) = A \left( a_{O}^2 + \sum_{l} \mu_l^2 f_{\textrm{bdry}} (\hat{\Delta}_l, \xi) \right)
\end{equation} 
where $A$ is the normalization of the operator. The blocks are known to be 
\begin{equation}
\begin{split}
f_{\textrm{bulk}} (\Delta_k ; \zeta) &= \left(\frac{\zeta}{4} \right)^{\frac{\Delta_k}{2}} \ \   _2F_1 \bigg(\frac{\Delta_k}{2},\frac{\Delta_k}{2}; \Delta_k + 1 - \frac{d}{2} ; -\frac{\zeta}{4} \bigg) \\
f_{\textrm{bdry}} ({\hat{\Delta}}_l;\zeta) &= \ \left(\frac{4}{\zeta} \right)^{{\hat{\Delta}}_l} \ \  _2 F_1 \bigg( {\hat{\Delta}}_l, {\hat{\Delta}}_l + 1 - \frac{d}{2}; 2 {\hat{\Delta}}_l + 2 - d; -\frac{4}{\zeta}  \bigg) .
\end{split}
\end{equation}
Expanding the two-point function \eqref{SigmaCorrMainB1} in powers of $1/\zeta$ tells us that the boundary spectrum consists of operators of dimension $d + 2 n$ with OPE coefficients given by
\begin{equation}
\mu^2_{d + 2 n} = \frac{\sqrt{\pi } 2^{-2 d-4 n+1} \Gamma \left(-\frac{d}{2}+n+2\right) \Gamma \left(\frac{d+1}{2}+n\right) \Gamma (d+2 n)}{\Gamma \left(2-\frac{d}{2}\right) \Gamma (n+1) \Gamma \left(d+n-\frac{1}{2}\right) \Gamma \left(\frac{d+1}{2} + 2n\right)}.
\end{equation}
The $n=0$ operator with $\hat\Delta_{(0)}=d$ corresponds to the displacement operator. 
In $d = 3$, the OPE coefficients simplify to \footnote{It does not quite agree with the result in \cite{Carmi:2018qzm}. We suspect this may be due to a different definition of the coefficients, or possibly a typo.}
\begin{equation}
\mu^2_{3 + 2 n} = \frac{(n + 1)^2}{4^{1 + 2 n} (4 n (2 + n) + 3)}.
\end{equation}
Note that there is no relevant scalar in the boundary theory in this phase. Hence this phase is the most stable one in the RG sense and must be at the end of the boundary RG flow, consistent with what we wrote in the introduction. In the bulk channel, the operators that appear are even powers of $\sigma$, i.e. $\sigma^{2k}$ with dimensions $2k$. The two-point function in the bulk OPE limit goes like
\begin{equation}
\langle \sigma (x_1) \sigma(x_2) \rangle = (\sigma^*)^2 - \frac{2^d \sin \left(\frac{\pi  d}{2}\right) \Gamma \left(\frac{d-1}{2}\right)}{\pi ^{3/2} c_d  N \Gamma \left(\frac{d}{2}-1\right) \zeta } - \frac{\Gamma (d+1)}{ c_d  N \Gamma \left(\frac{d}{2}-1\right)^2 \Gamma \left(1-\frac{d}{2}\right) \Gamma \left(\frac{d}{2}+1\right)}  \log \zeta + ...
\end{equation}
where the subleading terms are suppressed in the $\zeta \rightarrow 0$ limit. The $\log \zeta$ terms appear because the $\sigma^2$ operator already appears at leading order in $N$. So at order $1/N$, we expect the anomalous dimension of $\sigma^2$ to appear, which gives rise to the logarithm. From the structure of the OPE, the coefficient of the $\log$ should be related to the anomalous dimension as follows 
\begin{equation} \label{BulkOPESigmaLog}
(\sigma^*)^2 \left( \frac{\gamma_{\sigma^2}}{2} - \gamma_{\sigma} \right) = - \frac{\Gamma (d+1)}{ c_d  N \Gamma \left(\frac{d}{2}-1\right)^2 \Gamma \left(1-\frac{d}{2}\right) \Gamma \left(\frac{d}{2}+1\right)}.
\end{equation}
The bulk anomalous dimensions of $\sigma^2$  and $\sigma$ operators for the large N Gross-Neveu model are known \cite{Fei:2016sgs, Gracey:1990wi, Gracey:1992cp, Gracey:1993kb} and they satisfy the above relation, providing a non-trivial check of our results.

For $|\sigma^*| = d /2 - 2$, as we explain in appendix \ref{AppendixSigmaProp}, we have two choices for the $\sigma$ propagator. The first one has the following correlator \eqref{SigmaCorrAppP21}
\begin{equation}
\begin{split}
&G^D_{\sigma} (\zeta) = -\mathcal{B} \bigg[ \frac{2^{2-d}  \cos \left(\frac{\pi  d}{2}\right)}{(d-5) \Gamma \left(\frac{d-1}{2}\right)} \frac{1}{(4 + \zeta)^{2}} {}_2 F_1 \left(2, 3 - \frac{ d}{2}, 6 - d , \frac{4}{4 + \zeta}  \right) \\
&+ \frac{\Gamma(d)}{3 \Gamma \left(d - \frac{3}{2} \right) (d - 2)  \Gamma \left(\frac{d}{2}  - 2\right)  } \frac{\zeta + 2}{(\zeta(4 + \zeta))^{\frac{d + 1}{2}}} {}_3 F_2 \left( \frac{d + 1}{2}, \frac{d - 1}{2}, \frac{3}{2}; d - \frac{3}{2},  \frac{5}{2};  -\frac{4}{\zeta(4 + \zeta)}\right) \bigg]
\end{split}
\end{equation}
where $\mathcal{B}$ is a dimension dependent constant defined in \eqref{DefnB}. The boundary spectrum in this phase consists of a leading scalar of dimension $2$ and then a tower of operators of dimension $d + 2 n$ with the following OPE coefficients (the $n=0$ member of this tower should be, as above, the displacement operator)
\begin{equation} \label{BulkBounSigB2}
\begin{split}
\hat{\Delta}_l = \{ 2, d, d + 2, d + 4, d + 6,... \}, \hspace{1 cm}  \mu_2^2 =  -\frac{\sqrt{\pi } 2^{1-d} \cos \left(\frac{\pi  d}{2}\right) \Gamma \left(\frac{d}{2}-1\right)}{(d-5) \Gamma \left(\frac{d-1}{2}\right)} \\
\mu_{d + 2n}^2 = \frac{\sqrt{\pi } (-1)^{n+1} 2^{3-2 (d+2 n)} (d+2 n-1) \Gamma \left(\frac{d}{2}-1\right) \Gamma \left(\frac{d+1}{2}+n\right) \Gamma (d+2 n-2)}{(2 n+3) \Gamma (n+1) \Gamma \left(\frac{d}{2}-n-2\right) \Gamma \left(d+n-\frac{3}{2}\right)\Gamma \left(\frac{d+1}{2} + 2 n\right)}.
\end{split}
\end{equation}
This is the phase we called $B_2$ in the introduction and we use superscript $D$ to indicate that it matches on to GNY model with Dirichlet boundary condition on the scalar $s$. The dimension $2$ scalar operator we find here is relevant for $d > 3$ and may be turned on to flow to the $B_1$ phase. 

The $\sigma$ propagator for the $B_2'$ phase is \eqref{SigmaCorrAppP22}
\begin{equation}
\begin{split}
&G^N_{\sigma} (\zeta) = -\mathcal{B} \bigg[ -\frac{\pi ^{\frac{1}{2}} \Gamma \left(\frac{d}{2}-1\right)}{8 \Gamma \left(\frac{d-3}{2}\right) \Gamma \left(\frac{d-1}{2}\right)} \frac{1}{(4 + \zeta)^{d-3}} {}_2 F_1 \left(d-3, \frac{d}{2} - 2, d - 4 , \frac{4}{4 + \zeta}  \right) \\
&+ \frac{\Gamma(d)}{3 \Gamma \left(d - \frac{3}{2} \right) (d - 2)  \Gamma \left(\frac{d}{2}  - 2\right)  } \frac{\zeta + 2}{(\zeta(4 + \zeta))^{\frac{d + 1}{2}}} {}_3 F_2 \left( \frac{d + 1}{2}, \frac{d - 1}{2}, \frac{3}{2}; d - \frac{3}{2},  \frac{5}{2};  -\frac{4}{\zeta(4 + \zeta)}\right) \bigg].
\end{split}
\end{equation}
The boundary spectrum and the bulk-boundary OPE coefficients are the same as the $B_2$ phase, apart from the leading boundary scalar, which had dimensions $d - 3$ instead of $2$ and the OPE coefficient 
\begin{equation} \label{BulkBounSigB2'}
\mu^2_{d-3} = \frac{\pi  4^{3-d} \Gamma \left(\frac{d}{2}-1\right)^2}{\Gamma \left(\frac{d-3}{2}\right) \Gamma \left(\frac{d-1}{2}\right)}.
\end{equation}  
The relevant operator $\hat{\sigma}^2$ of dimensions $2 d - 6$ drives the flow from the $B_2'$ to $B_2$ phase. The bulk spectrum in $B_2$ and $B_2'$ phases is of course the same as in $B_1$ phase. The two-point function in the bulk OPE limit still contains a $\log \zeta$ whose coefficient is related to the bulk anomalous dimension of the $\sigma^2$ operator, as we saw in \eqref{BulkOPESigmaLog} for the $B_1$ phase. We now expand these propagators in $d = 4 - \epsilon$
\begin{equation} \label{SigmaCorrEps}
\begin{split}
G_{\sigma}^{N} &= \frac{\epsilon}{N} \left( \frac{1}{\zeta} + \frac{1}{4 + \zeta} \right) - \frac{  \epsilon^2}{N}  \left( \frac{1}{\zeta} - \frac{1}{4 + \zeta} \right)    + \frac{\epsilon^2}{N} \left( \frac{1}{\zeta} + \frac{1}{4 + \zeta}  \right) \log \left( 1 +  \frac{\zeta}{4} \right) +  O(\epsilon^3) \\
G_{\sigma}^{D} &= \frac{\epsilon - \epsilon^2}{N} \left( \frac{1}{\zeta} - \frac{1}{4 + \zeta} \right) + O(\epsilon^3).
\end{split}
\end{equation}
As we will see in section \ref{Sec:EOM}, these exactly match the correlator of $s$ in GNY model, once we normalize the operators in the same way. Note that in $d=3$ the dimension of the leading boundary scalar induced by $\sigma$ becomes zero at large $N$. This may indicate that the $B_2'$ boundary conformal phase may not survive in $d=3$, though it is present in the range $3<d<4$. It would be interesting to clarify this. 

\subsection{Free Energy}
In this subsection, we calculate the AdS free energy at the large $N$ boundary fixed points we discussed. At leading order, the one-point function of $\sigma$ acts as a mass for fermions, so we can use the results from section \ref{Sec:FreeMassiveFermion}. For $d = 3$, we can just use the general formula \eqref{FreeEnergy3dGen}. For the two phases, we get 
\begin{equation}
F (3/2) = - \frac{N \mathrm{Vol} (H^3) }{24 \pi}, \hspace{1cm} F (1/2) = \frac{N \mathrm{Vol} (H^3) }{24 \pi}.
\end{equation}
The value of the trace anomaly coefficient for these phases is  \eqref{a3dDef}
\begin{equation}
a_{3d} (3/2) = - \frac{N }{24 }, \hspace{1cm} a_{3d}(1/2) = \frac{N }{24 }.
\end{equation}

\begin{figure} 
\centering
\includegraphics[scale=1]{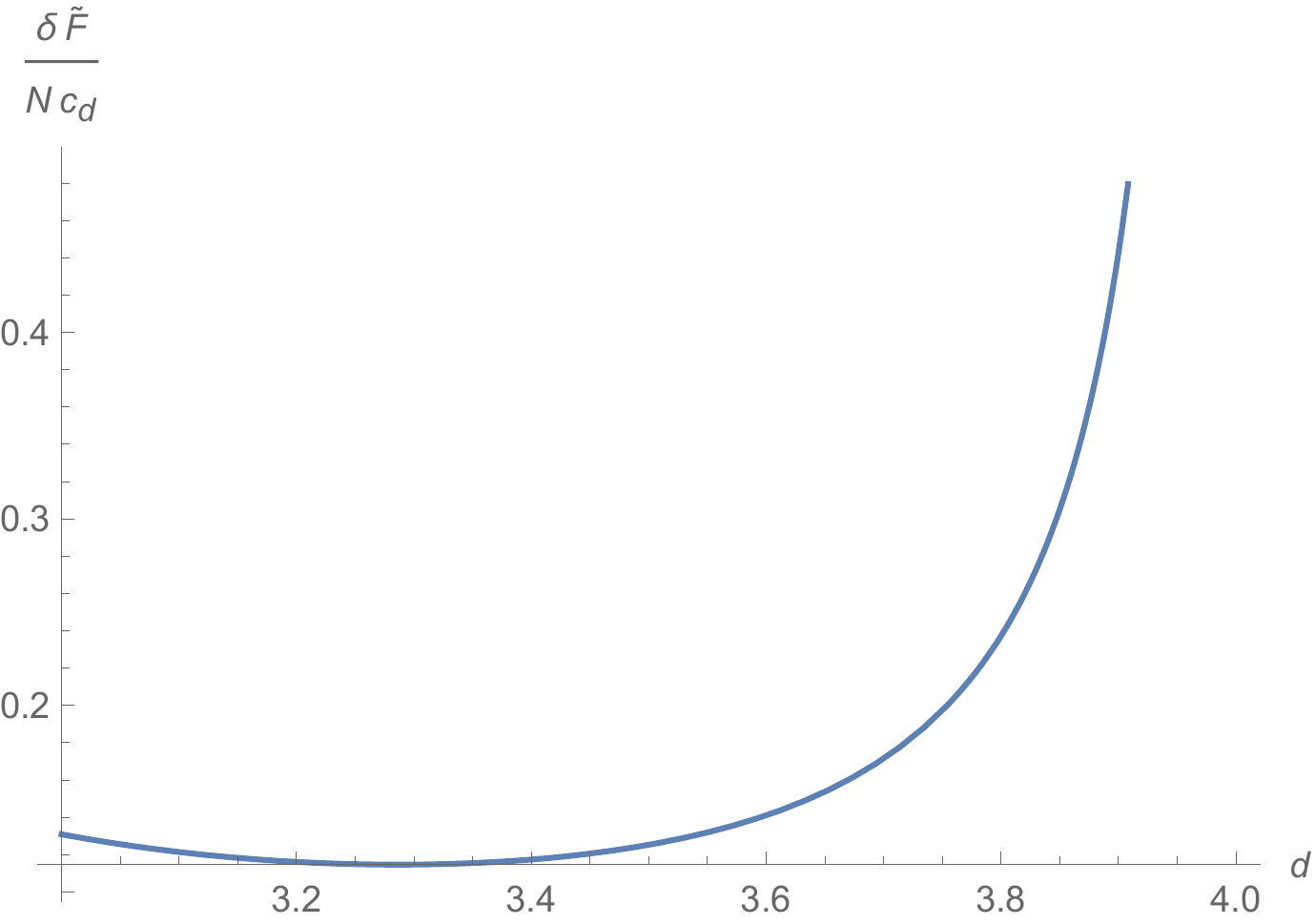}
\caption{The free energy difference at large $N$ between the two phases, $\delta \tilde{F} = \tilde{F}_{\sigma^* = 2 -d/2 } - \tilde{F}_{\sigma^* = d/2 - 1}$ where both the phases exist, i.e. between $3 < d < 4$.}
\label{FigureLargeNFDiff}
\end{figure}

For other values of $d$, the free energy can be calculated in terms of some reference value, say the free energy of free massless fermions. For the $B_1$ phase, using \eqref{FreeEneryDer2}, we have 
\begin{equation} 
\begin{split}
F_{\sigma^* = d/2 - 1} &= F[\sigma^* = 0] \ + \ \int_{0}^{d/2 - 1} d \sigma \frac{\partial F}{\partial \sigma} \\
&= N F_{\textrm{Free}} \ - \ \frac{ N c_{d} \mathrm{Vol} (H^d) \Gamma(1-\frac{d}{2}) }{ (4 \pi)^{\frac{d}{2}} }  \int_{0}^{d/2 - 1} d \sigma \frac{  \Gamma  \left(\frac{d}{2}+ \sigma \right) }{ \Gamma \left(1 - \frac{d}{2} + \sigma \right)}
\end{split}
\end{equation}
where $F_{\textrm{Free}}$ is the free energy of a single free massless fermion on $H^d$. It is easy to see that the free energy itself does not depend on the sign of $\sigma^*$, so we restrict ourselves to positive $\sigma^*$ in this section. In $d = 2 + \epsilon$, this has the following expression
\begin{equation} \label{Phase1Two}
F_{\sigma^* = d/2 - 1} \big|_{d = 2 + \epsilon} = N F_{\textrm{Free}}  -\frac{ N \mathrm{Vol} (H^2)}{8 \pi}  \epsilon + O(\epsilon^2)
\end{equation}
while in $d = 4 - \epsilon$, this gives 
\begin{equation} \label{Phase1Four}
F_{\sigma^* = d/2 - 1} \big|_{d = 4 - \epsilon} = N F_{\textrm{Free}} \ - \ \frac{ N \mathrm{Vol} (H^4)}{8 \pi^2 \epsilon}   + O(1).
\end{equation}
For $\sigma^* = 2 - d/2 $, using \eqref{FreeEneryDer1} we have 
\begin{equation} 
\begin{split}
F_{\sigma^* = 2 -d/2 } &= F[\sigma^* = 0] \ + \ \int_{0}^{2 - d/2} d \sigma \frac{\partial F}{\partial \sigma} \\
&= N F_{\textrm{Free}} \ + \ \frac{ N c_{d} \mathrm{Vol} (H^d) \Gamma(1-\frac{d}{2}) }{ (4 \pi)^{\frac{d}{2}} }  \int_{0}^{2 - d/2} d \sigma \frac{  \Gamma  \left(\frac{d}{2} - \sigma \right) }{ \Gamma \left(1 - \frac{d}{2} - \sigma \right)}
\end{split}
\end{equation}
In $d = 4 - \epsilon$, this is 
\begin{equation} \label{Phase2Four}
F_{\sigma^* = 2- d/2} \big|_{d = 4 - \epsilon} = N F_{\textrm{Free}} \ + \ \frac{ N \mathrm{Vol} (H^4) \epsilon}{16 \pi^2 }.
\end{equation}
In the next section, we will match these with the calculation in $\epsilon$ expansion. As we mentioned in the introduction, we expect a RG flow from $B_2$ to $B_1$ phase, so we expect $\tilde{F}$ defined in \eqref{FTildeDefn} to be lower for the $B_1$ phase. It can be seen numerically that $\tilde{F}$ for $\sigma^* = d/2 - 1$ is lower than that for $\sigma^* = 2 - d/2$. We plot the difference in $\tilde{F}$ between these phases in Figure \ref{FigureLargeNFDiff}. 

There is also an RG flow from $B_2'$ to $B_2$ phase, so we also expect $\tilde{F}$ for $B_2$ phase to be lower than that for $B_2'$ phase. To calculate the free energy difference, we can think of the flow between the two as a double trace flow on the boundary triggered by a $\hat{\sigma}^2$ operator. The free energy change under an RG flow driven by the square of a scalar operator in a large $N$ CFT is given by \cite{Diaz:2007an, Giombi:2014xxa, Giombi:2020rmc} 
\begin{equation} \label{DoubleTraceScalar}
F_{d - 1- \hat{\Delta}} - F_{\hat{\Delta}} = - \frac{1}{\sin \left( \frac{\pi (d - 1)}{2} \right) \Gamma(d)} \int_0^{\hat{\Delta} - \frac{d - 1}{2}} d u u \sin \pi u \Gamma\left( \frac{d - 1}{2} + u \right) \Gamma\left( \frac{d - 1}{2} - u \right).
\end{equation}
Applying it for $\hat{\Delta} = 2$, we get 
\begin{equation}
F^N - F^D =  F_{d - 3} - F_{2} = - \frac{1}{\sin \left( \frac{\pi (d - 1)}{2} \right) \Gamma(d)} \int_0^{\frac{5 - d}{2}} d u u \sin \pi u \Gamma\left( \frac{d - 1}{2} + u \right) \Gamma\left( \frac{d - 1}{2} - u \right).
\end{equation}
We can use this to calculate the difference in $\tilde{F} = - \sin \left( \frac{\pi (d - 1)}{2} \right) F$ and check that $\tilde{F}_{d - 3} - \tilde{F}_{2}$ is positive between $3 < d < 4$. In $4 - \epsilon$, we get 
\begin{equation} \label{LargeNNeumannDirichlet}
\begin{split}
F^N - F^D &=  - \frac{1}{\sin \left( \frac{\pi (d - 1)}{2} \right) \Gamma(d)} \int_0^{\frac{1}{2}} d u u \sin \pi u \Gamma\left( \frac{d - 1}{2} + u \right) \Gamma\left( \frac{d - 1}{2} - u \right) + \frac{\epsilon}{24} + O(\epsilon^2)\\
  &= \frac{\zeta(3)}{8 \pi^2} + A \epsilon + O(\epsilon^2)
\end{split}
\end{equation}
and numerically, $A = 0.06122$. We will check this against an $\epsilon$ expansion calculation in GNY model in next section. 

\section{$\epsilon$ expansion} \label{Sec:EpsExp}
In this section, we study alternative descriptions of the above fixed points near $d = 2$ and $d = 4$. These results are valid for all $N$.

\subsection{Gross-Neveu model in $d = 2 + \epsilon$}
We start with Gross-Neveu model described in \eqref{ActionGN} and study it near two dimensions, where there is only one boundary phase $B_1$. There is a fixed point at 
\begin{equation}
g^* = \frac{ \pi}{N - 1} \epsilon.
\end{equation}
The $\sigma$ operator in the large $N$ theory is related, by equation of motion, to $ \bar{\Psi}_i \Psi^i$ as
\begin{equation}
\sigma^* = g^* \langle \bar{\Psi}_i \Psi^i \rangle.
\end{equation}
Taking the short distance limit of \eqref{MasslessProp1}, we get
\begin{equation} \label{PsibarPsiOnePoint}
\langle \bar{\Psi} \Psi \rangle = \mp \frac{ c_d \Gamma\left(\frac{d}{2} \right)}{ (4 \pi)^{\frac{d}{2}} } \implies \sigma^* =  \mp \frac{N}{2 (N - 1)} \epsilon.
\end{equation}
At large $N$, this agrees with \eqref{LargeNDimPhase1}. The two possible signs of $\sigma^*$ correspond to two different boundary conditions we can impose on the fermion, and define two equivalent theories. The free energy to leading order in $\epsilon$  in $d = 2 + \epsilon$ is given by

\begin{equation}
\begin{split}
F & = N F_{\textrm{Free}} - \frac{g^*}{2} \textrm{Vol}(H^d) N \left( N - \frac{1}{2} \right) \langle \bar{\Psi} \Psi \rangle^2   \\
&= N F_{\textrm{Free}} - \frac{\textrm{Vol}(H^2) N (2 N - 1)}{16 \pi (N - 1) } \epsilon 
\end{split}
\end{equation}
At large $N$, this matches the large $N$ result in \eqref{Phase1Two}.

Let us now look at the spectrum of the boundary theory. One way to do this is to calculate the boundary correlation functions in the $\epsilon$ expansion. We will just do the calculation for the $+$ boundary condition on the fermion, but it goes exactly the same way for the other case. Let's start with the two-point function. In the free theory, it is given by \eqref{BoundaryRepTwoP} with $\hat{\Delta} = (d-1)/2$
\begin{equation} 
\langle \psi^i (\textbf{x}_1) \bar{\psi}_j (\textbf{x}_2) \rangle_0 =  -\frac{\Gamma \left(\frac{d}{2} \right)}{ \pi^{\frac{d}{2}}} \frac{ \delta^i_j  \boldsymbol{\Gamma} \cdot \textbf{x}_{12}}{(\textbf{x}_{12}^2)^{\frac{d}{2}}}.
\end{equation}
In the interacting Gross-Neveu model, the two point function receives corrections which can be calculated using the bulk tadpole Witten diagram. We will need bulk-boundary propagator \eqref{FermionBulkBoundaryGen}, so we will calculate the interaction piece first when the fermions are in the bulk spinor representation, and then project onto the boundary representation
\begin{equation}
\begin{split}
\langle \hat{\Psi}^{i} (\textbf{x}_{1})  \hat{\bar{\Psi}}_{j} (\textbf{x}_{2}) \rangle_1 &= \ \vcenter{\hbox{\includegraphics[scale=0.3]{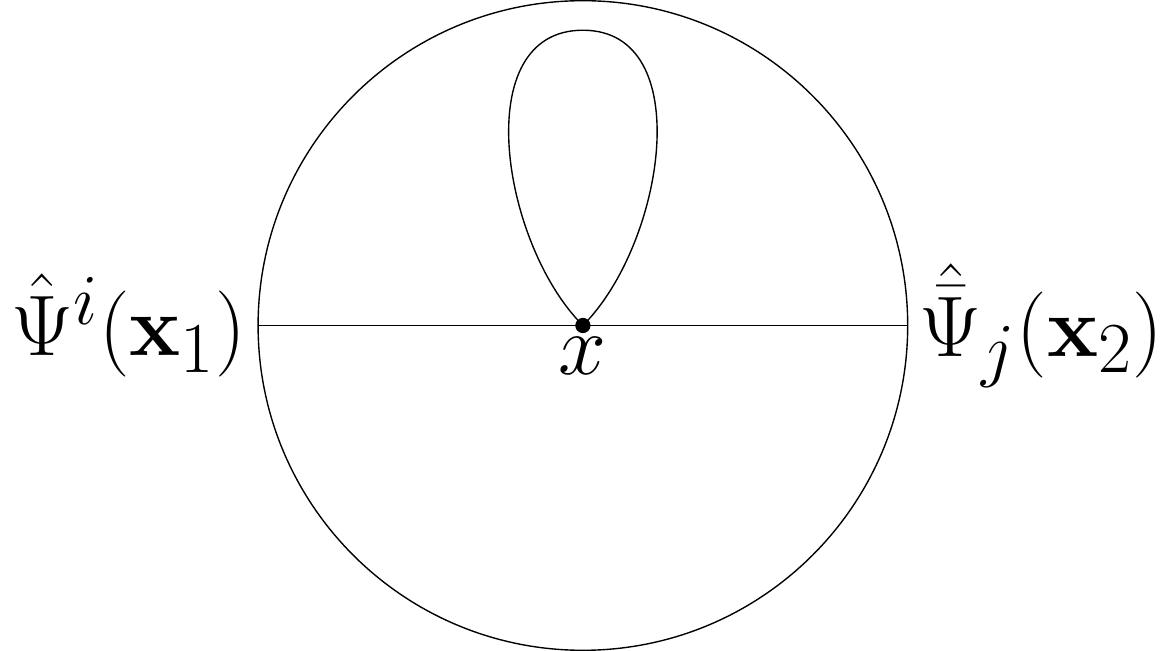}}} \\ 
& = -g \delta^i_j \left( N - \frac{1}{c_d} \right) \langle \bar{\Psi} \Psi \rangle \int d^d x \sqrt{g_x}  \langle \hat{\Psi} (\textbf{x}_1) \bar{\Psi} (x) \rangle \langle \hat{\bar{\Psi}} (\textbf{x}_2) \Psi (x) \rangle.
\end{split}
\end{equation}
Using \eqref{FermionBulkBoundaryGen}, we note that the product of two bulk-boundary propagators for fermions can be simplified as 
\begin{equation} \label{BulkBoundaryProduct}
\begin{split}
\langle \hat{\Psi} (\textbf{x}_1) \bar{\Psi} (x) \rangle \langle \hat{\bar{\Psi}} (\textbf{x}_2) \Psi (x) \rangle &= \frac{\Gamma \left( \frac{d}{2} \right)^2}{4 \pi^d} \frac{z^{d - 1} (1 + \gamma_0) \gamma_a (x_1 - x)^a \gamma_b (x_2 - x)^b (1 - \gamma_0) }{\left((z^2 + (\textbf{x}_1 - \textbf{x})^2)(z^2 + (\textbf{x}_2 - \textbf{x})^2) \right)^{\frac{d}{2}}} \\
&= \frac{\Gamma \left( \frac{d}{2} \right)^2}{2 \pi^d} \frac{z^{d} (1 + \gamma_0) \boldsymbol{\gamma} \cdot \textbf{x}_{12}}{\left((z^2 + (\textbf{x}_1 - \textbf{x})^2)(z^2 + (\textbf{x}_2 - \textbf{x})^2) \right)^{\frac{d}{2}}}
\end{split}
\end{equation}
Plugging in all the factors near $d = 2$ gives, to leading order in $\epsilon$
\begin{equation}
\begin{split}
\langle \hat{\Psi}^{i} (\textbf{x}_{1})  \hat{\bar{\Psi}}_{j} (\textbf{x}_{2}) \rangle_1 &= \frac{\delta^i_j \epsilon (2 N - 1) (1 + \gamma_0) \boldsymbol{\gamma} \cdot \textbf{x}_{12} }{8 \pi^2 (N - 1)}  \int d^d x \frac{1}{{\left((z^2 + (\textbf{x}_1 - \textbf{x})^2)(z^2 + (\textbf{x}_2 - \textbf{x})^2) \right)^{\frac{d}{2}}}} \\
&= \frac{\delta^i_j \epsilon (2 N - 1) (1 + \gamma_0) \boldsymbol{\gamma} \cdot \textbf{x}_{12} }{16 \pi (N - 1)}  \int d \alpha  d z \frac{1}{(z^2 + \alpha (1 - \alpha) \textbf{x}_{12}^2)^{\frac{3}{2}}} .
\end{split}
\end{equation}
The integral above has a divergence at $z = 0$, and it  corresponds to an anomalous dimension for the boundary operator $\hat{\Psi}$. To calculate this anomalous dimension, we only need the logarithmic piece of the above integral, which can be extracted by regulating it as follows 
\begin{equation}
\begin{split}
\langle \hat{\Psi}^{i} (\textbf{x}_{1})  \hat{\bar{\Psi}}_{j} (\textbf{x}_{2}) \rangle_1 &= \frac{\delta^i_j \epsilon (2 N - 1) (1 + \gamma_0) \boldsymbol{\gamma} \cdot \textbf{x}_{12} }{16 \pi (N - 1)}  \int d \alpha  d z \frac{z^{\eta}}{(z^2 + \alpha (1 - \alpha) \textbf{x}_{12}^2)^{\frac{3}{2}}} \\
&= \frac{\delta^i_j \epsilon (2 N - 1) (1 + \gamma_0) \boldsymbol{\gamma} \cdot \textbf{x}_{12} }{8 \pi (N - 1) \textbf{x}_{12}^2} \left( \frac{2}{\eta} + \log \left( \textbf{x}_{12}^2 \right) - 2 \log 2 + O(\eta) \right).
\end{split}
\end{equation}
Projecting it onto the boundary gives 
\begin{equation}
\langle \psi^i (\textbf{x}_1) \bar{\psi}_j (\textbf{x}_2) \rangle_1 = \frac{\delta^i_j \epsilon (2 N - 1) \boldsymbol{\Gamma} \cdot \textbf{x}_{12} }{4 \pi (N - 1) \textbf{x}_{12}^2} \left( \frac{2}{\eta} + \log \left( \textbf{x}_{12}^2 \right) - 2 \log 2 + O(\eta) \right).  
\end{equation}
The $\log$ piece gives us the anomalous dimension of the leading boundary fermion 
\begin{equation} \label{BoundaryFAnomGN}
\hat{\gamma} = \frac{(2 N - 1)}{4 (N - 1)} \epsilon, \implies \hat{\Delta} = \frac{d-1}{2} + \hat{\gamma} = \frac{1}{2} + \frac{4 N - 3}{4 (N - 1)} \epsilon.
\end{equation}
This is consistent with the large $N$ result of $d - 3/2$ \eqref{LargeNDimPhase1}.

Next, let us calculate the four-point function on the boundary. This should give anomalous dimensions of the scalar operators on the boundary, which are bilinears of the leading fermionic operator. In the free theory, the four-point function is given by Wick contractions of \eqref{BoundaryRepTwoP}
\begin{equation}
\begin{split}
&\langle \bar{\psi}_{i, a} (\textbf{x}_{1}) \psi^{j, b} (\textbf{x}_{2})  \bar{\psi}_{k, c} (\textbf{x}_{3})  \psi^{l, d} (\textbf{x}_{4})\rangle_0 = \\
&\frac{\Gamma \left(\frac{d}{2} \right)^2 }{\pi^d} \left( \frac{ \delta_i^j \delta_k^l {\left(\boldsymbol{\Gamma} \cdot \textbf{x}_{12} \right)^b}_{a}  { \left(\boldsymbol{\Gamma} \cdot \textbf{x}_{12} \right)^d}_{c}}{(\textbf{x}_{12}^2)^{\frac{d}{2}}(\textbf{x}_{34}^2)^{\frac{d}{2}}} + \frac{ \delta_i^l \delta_k^j {\left(\boldsymbol{\Gamma} \cdot \textbf{x}_{14} \right)^d}_{a}  { \left(\boldsymbol{\Gamma} \cdot \textbf{x}_{23} \right)^b}_{c}}{(\textbf{x}_{14}^2)^{\frac{d}{2}}(\textbf{x}_{23}^2)^{\frac{d}{2}}}  \right) 
\end{split}
\end{equation}
where indices $a, b, c , d$ are boundary spinor indices. We now restrict to two bulk dimensions, so that boundary is one-dimensional and the boundary gamma matrix is just $1$ 
\begin{equation}
\begin{split}
\langle \bar{\psi}_i (\textbf{x}_{1}) \psi^{j} (\textbf{x}_{2})  \bar{\psi}_{k} (\textbf{x}_{3})  \psi^{l} (\textbf{x}_{4})\rangle_0 &= \delta_{i}^j \delta_k^l \mathcal{S}_{\textrm{sing}} + \left( \delta_{i}^l \delta_k^j - \frac{\delta_{i}^j \delta_k^l }{N}  \right)\mathcal{S}_{\textrm{adj}}   \\
&=  \frac{\Gamma \left(\frac{d}{2} \right)^2 }{\pi^d} \left( \frac{ \delta_i^j \delta_k^l  \textbf{x}_{12} \textbf{x}_{34}}{(\textbf{x}_{12}^2)^{\frac{d}{2}}(\textbf{x}_{34}^2)^{\frac{d}{2}}} + \frac{ \delta_i^l \delta_k^j \textbf{x}_{14}\textbf{x}_{23} }{(\textbf{x}_{14}^2)^{\frac{d}{2}}(\textbf{x}_{23}^2)^{\frac{d}{2}}}  \right) 
\end{split}
\end{equation}
where we defined the $U(N)$ singlet and adjoint parts of the four-point function. For convenience, we now restrict to the configuration $\textbf{x}_1 > \textbf{x}_2 > \textbf{x}_3 > \textbf{x}_4$. The first term in the correlator above represents the contribution of the identity operator, while the second term contains contributions of operators appearing in the OPE of $\bar{\psi}(x_1) \psi(x_2)$ and can be decomposed into conformal blocks using \cite{Carmi:2018qzm}
\begin{equation}
\frac{1}{ \left( \textbf{x}_{14} \textbf{x}_{14} \right)^{2 \hat{\Delta}}} = \frac{1}{ \left( \textbf{x}_{12} \textbf{x}_{34} \right)^{2 \hat{\Delta}}} \sum_{n = 0}^{\infty} c_n^2 \mathcal{K}_{\hat{\Delta}_n} (\chi) , \hspace{1cm} \hat{\Delta}_n = 2 \hat{\Delta} + n, \hspace{1cm} \chi = \frac{\textbf{x}_{12} \textbf{x}_{34}}{\textbf{x}_{13} \textbf{x}_{24}}. 
\end{equation}
The intermediate scalar operators have the schematic form $\bar{\psi} \left( \boldsymbol{\slashed{\partial}} \right)^n \psi $ and the OPE coefficients and conformal blocks turn out to be \cite{Carmi:2018qzm}
\begin{equation}
c_n^2 =  \frac{4^{1 - n} \hat{\Delta} \ (4 \hat{\Delta})_{n-1} (2 \hat{\Delta} + 1)_{n-1}}{n! \left( 2 \hat{\Delta} + \frac{1}{2} \right)_{n-1}}, \hspace{1cm} \mathcal{K}_{\hat{\Delta}_n} (\chi) = \chi^{\hat{\Delta}_n} {}_2 F_1 \left( \hat{\Delta}_n, \hat{\Delta}_n, 2 \hat{\Delta}_n, \chi \right).
\end{equation}
In the free theory, the dimensions of composite operators are just $\hat{\Delta}_n = (d - 1) + n$, but in the interacting theory, they get corrected to $\hat{\Delta}_n = 2 \hat{\Delta} + n + \hat{\gamma}_n$.

The interaction term corrects this four-point function, which can be calculated using the bulk contact Witten diagram
\begin{equation} 
\begin{split}
\langle \hat{\bar{\Psi}}_{i, \alpha} (\textbf{x}_{1})  \hat{\Psi}^{j, \beta} (\textbf{x}_{2})  \hat{\bar{\Psi}}_{k, \gamma} (\textbf{x}_{3})   \hat{\Psi}^{l, \delta} (\textbf{x}_{4})\rangle_1 = \ \vcenter{\hbox{\includegraphics[scale=0.3]{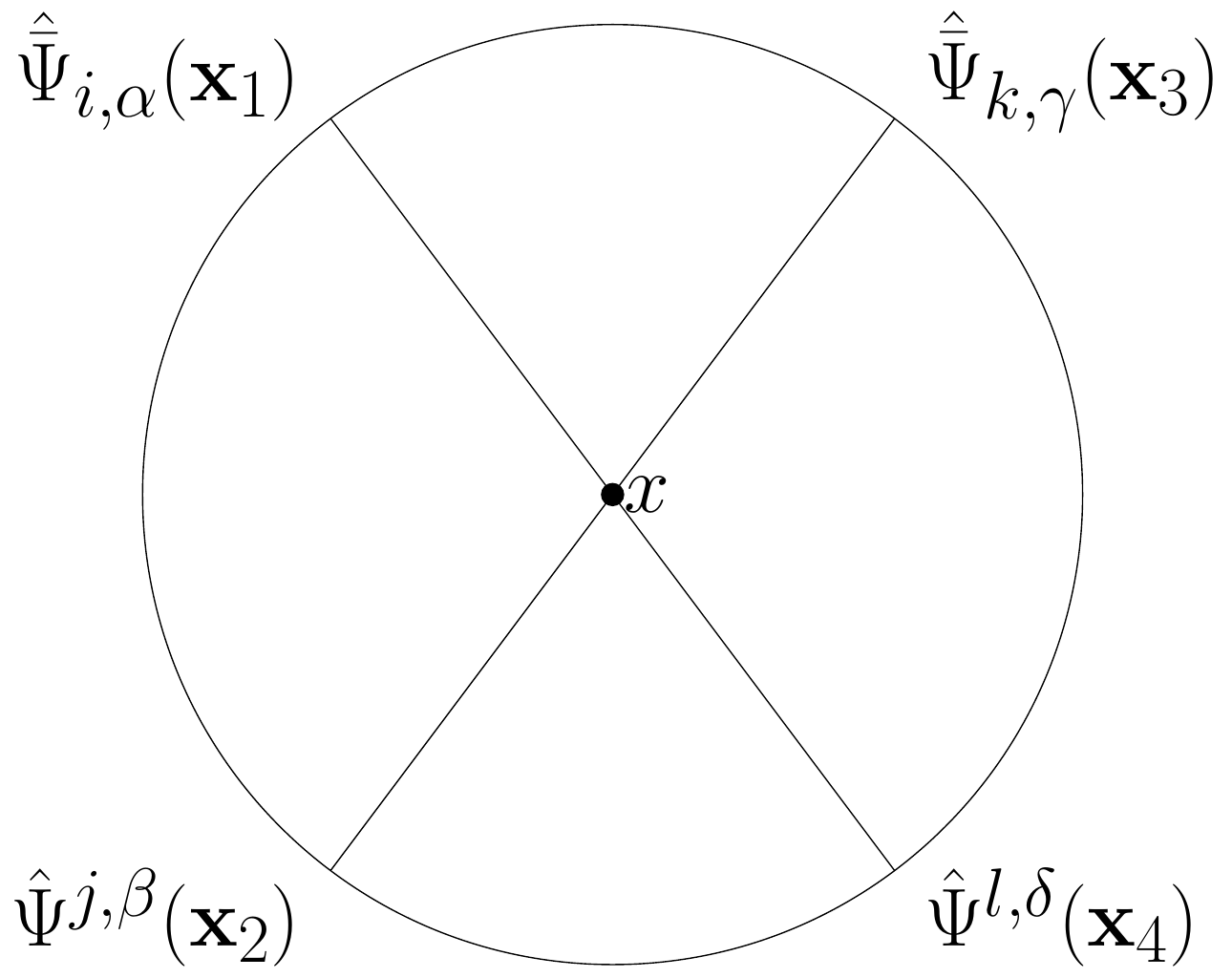}}} \\
= \frac{g}{2} \int d^dx \sqrt{g_x} \langle \hat{\bar{\Psi}}_{i, \alpha} (\textbf{x}_{1})  \hat{\Psi}^{j, \beta} (\textbf{x}_{2})  \hat{\bar{\Psi}}_{k, \gamma} (\textbf{x}_{3})   \hat{\Psi}^{l, \delta} (\textbf{x}_{4}) (\bar{\Psi}_m \Psi(x)^m)^2 \rangle. 
\end{split}
\end{equation}
Here, the greek indices $\alpha, ...$ are bulk spinor indices. Plugging in the bulk-boundary propagators and using \eqref{BulkBoundaryProduct} gives the following integral for the contact interaction 
\begin{equation}
\begin{split}
\langle \hat{\bar{\Psi}}_{i, \alpha} (\textbf{x}_{1})  \hat{\Psi}^{j, \beta} (\textbf{x}_{2})  \hat{\bar{\Psi}}_{k, \gamma} (\textbf{x}_{3})   \hat{\Psi}^{l, \delta} (\textbf{x}_{4})\rangle_1 = \frac{g \Gamma \left( \frac{d}{2} \right)^4  }{4 \pi^{2 d}} \int d^d x z^{-d} \prod_{m = 1}^4 \left( \frac{z}{z^2 + (\textbf{x}_m - \textbf{x})^2} \right)^{\frac{d}{2}} \times \\ 
\left[ \delta_i^j \delta_k^l {\left( (1 + \gamma_0) \boldsymbol{\gamma} \cdot \textbf{x}_{12}\right)^{\beta}}_{\alpha} {\left( (1 + \gamma_0) \boldsymbol{\gamma} \cdot \textbf{x}_{34}\right)^{\delta}}_{\gamma} + \delta_i^l \delta_k^j {\left( (1 + \gamma_0) \boldsymbol{\gamma} \cdot \textbf{x}_{14}\right)^{\delta}}_{\alpha} {\left( (1 + \gamma_0) \boldsymbol{\gamma} \cdot \textbf{x}_{23}\right)^{\beta}}_{\gamma}   \right].
\end{split}
\end{equation}
We can contract this with bulk polarization spinors \eqref{BoundaryPolarization} and then differentiate with respect to boundary polarization spinors exactly as we did for the two-point function, to get the four-point function in the boundary spinor representation. The integral can be evaluated in terms of well known $D$-functions \cite{DHoker:1999kzh, Dolan:2000ut} defined as the following AdS integral
\begin{equation}
D_{\Delta_1, \Delta_2, \Delta_3, \Delta_4} (\textbf{x}_1, \textbf{x}_2,\textbf{x}_3, \textbf{x}_4) = \int \frac{d^d x}{z^d} \prod_{i = 1}^4 \left( \frac{z}{z^2 + (\textbf{x} - \textbf{x}_i)^2} \right)^{\Delta_i}.
\end{equation}
In $d = 2$, the explicit expression for this $D$ function can be worked out in terms of elementary functions (see for example \cite{Giombi:2017cqn, Beccaria:2019dws}). This gives the correction to the four-point function 
\begin{equation} 
\begin{split}
\langle \bar{\psi}_i (\textbf{x}_{1}) \psi^{j} (\textbf{x}_{2})  \bar{\psi}_{k} (\textbf{x}_{3})  \psi^{l} (\textbf{x}_{4})\rangle_1 &=  \frac{ g \Gamma \left(\frac{d}{2} \right)^4 }{\pi^{2 d}} \left(  \delta_i^j \delta_k^l  \textbf{x}_{12} \textbf{x}_{34} + \delta_i^l \delta_k^j \textbf{x}_{14}\textbf{x}_{23} \right) D_{\frac{d}{2}, \frac{d}{2}, \frac{d}{2}, \frac{d}{2}} \\
&= - \frac{ \epsilon \left( \chi \ \delta_i^j \delta_k^l  + (1 - \chi) \ \delta_i^l \delta_k^j \right) \chi}{ 2 (N - 1)\pi^{2} \textbf{x}_{12} \textbf{x}_{34}} \left( \frac{\log \chi}{1 - \chi} + \frac{\log (1 - \chi)}{\chi} \right). 
\end{split}
\end{equation}
The $\log \chi$ piece above gives the anomalous dimensions of the composite operators. This is because in the conformal block decompostion, the $\log \chi$ comes from the derivative of the conformal block with respect to the dimension  
\begin{equation}
\partial_n \mathcal{K}_{\hat{\Delta}_n} (\chi) = \log( \chi) \chi^{\hat{\Delta}_n} {}_2 F_1 \left( \hat{\Delta}_n, \hat{\Delta}_n, 2 \hat{\Delta}_n, \chi \right).
\end{equation}
Hence the conformal block expansion contains 
\begin{equation}
\sum_{n = 0}^{\infty} c_n^2 \hat{\gamma}_n \log( \chi) \chi^{1 + n} {}_2 F_1 \left( 1 + n, 1 + n, 2 + 2 n, \chi \right). 
\end{equation}
Comparing the $\log$ terms, we get the following anomalous dimensions in the $U(N)$ adjoint sector 
\begin{equation}
\begin{split}
\sum_{n = 0}^{\infty} c_n^2 \hat{\gamma}^{\textrm{adj}}_n  \chi^{1 + n} {}_2 F_1 \left( 1 + n, 1 + n, 2 + 2 n, \chi \right) = -\frac{\epsilon \chi}{2 (N - 1)}   \\
\hat{\gamma}^{\textrm{adj}}_0 = -\frac{\epsilon}{2 (N - 1)} \implies \hat{\Delta}^{\textrm{adj}}_0 = 2 \hat{\Delta} + \hat{\gamma}^{\textrm{adj}}_0 = 1 + 2 \epsilon
\end{split}
\end{equation}
where we used the corrected dimension of the boundary fermion operator from \eqref{BoundaryFAnomGN}. The anomalous dimensions of all the other higher operators vanish. At large $N$, this matches $2 \hat{\Delta}$  resulting from the large N calculation \eqref{LargeNDimPhase1}. This is what we expect because in the large $N$ theory, the correction to $2 \hat{\Delta}$ comes from the connected $\sigma$ exchange diagram, which should be suppressed at large $N$ in the adjoint sector. Similarly, in the singlet sector, we get the following equation to determine anomalous dimension
\begin{equation}
\sum_{n = 0}^{\infty} c_n^2 \hat{\gamma}^{\textrm{sing}}_n \chi^{1 + n} {}_2 F_1 \left( 1 + n, 1 + n, 2 + 2 n, \chi \right) = -\frac{\epsilon \chi}{2 (N - 1)} \left(1 + \frac{N \chi}{1 - \chi} \right).
\end{equation}
Expanding both sides in powers of $\chi$ gives anomalous dimensions of all the fermion bilinears in the OPE $\bar{\psi}$ and $\psi$. We just write the dimensions of the first two operators 
\begin{equation}
\begin{split}
&\hat{\gamma}^{\textrm{sing}}_0 = -\frac{\epsilon}{2 (N - 1)} \implies \hat{\Delta}^{\textrm{sing}}_0 = 2 \hat{\Delta} + \hat{\gamma}^{\textrm{sing}}_0 = 1 + 2 \epsilon. \\
&\hat{\gamma}^{\textrm{sing}}_1 = -\frac{\epsilon (2 N - 1)}{2 (N - 1)} \implies \hat{\Delta}^{\textrm{sing}}_1 = 2 \hat{\Delta} + 1 + \hat{\gamma}^{\textrm{sing}}_1 = 2 +  \epsilon
\end{split}
\end{equation}
where again, we used the corrected dimension of the boundary fermion operator from \eqref{BoundaryFAnomGN}. The $n = 0$ operator is the leading singlet scalar operator on the boundary. The $n = 1$ operator is proportional to the displacement operator and has dimension $d = 2 + \epsilon$. We expect this dimension to stay protected to all orders in the perturbation theory. Also, this $n = 1$ singlet operator is the one that in the large $N$ theory corresponds to $\hat{\sigma}$ and its dimension was referred to as $\hat{\Delta}_{(0)}$ in Table \ref{TableBoundaryPhasesDimensions}.

\subsection{Gross-Neveu-Yukawa model in $d = 4 - \epsilon$}
In $d = 4 - \epsilon$, the large $N$ theory should match the Gross-Neveu-Yukawa model, which in hyperbolic space, may be described by the following action 
\begin{equation} \label{ActionGNYAdS}
S = \int d^d x \sqrt{g(x)} \left( \frac{(\partial_{\mu} s)^2}{2} - \frac{d (d-2)}{8} s^2 - \left( \bar{\Psi}_i \gamma\cdot \nabla \Psi^i + g_1 s \bar{\Psi}_i \Psi^i\right) + \frac{g_2}{24} s^4   \right)
\end{equation}
where $i = 1, ..., N$, so we have $N$ Dirac fermions. There is a fixed point at the following values of the couplings 
\begin{equation}
\begin{split}
(g_1^*)^2 &= \frac{(4\pi)^2}{4 N + 6 } \epsilon \xrightarrow{\textrm{Large N}} 
\frac{4\pi^2}{N} \epsilon \\
g_2^* &= \frac{(4\pi)^2 \left( - 2 N + 3  + \sqrt{4 N ^2 + 132 N + 9} \right)}{3 (4 N + 6)} \epsilon \xrightarrow{\textrm{Large N}}
 \frac{3 (4\pi)^2}{N} \epsilon.
\end{split}
\end{equation}
The operator $s$ in this description is proportional to the $\sigma$ operator in the large $N$ description. In the $B_1$ phase, $s$ gets a vev in the classical theory. It appears naturally in the hyperbolic space as the minimum of the potential which occurs at
\begin{equation} \label{GNYVev}
(s^*)^2 =   \frac{3 d (d-2)}{2 g_2^*} \implies |\sigma^*| = |g_1^* s^*| =  \sqrt{\frac{36}{-2 N + 3 + \sqrt{4 N ^2 + 132 N + 9}}} + O(\epsilon)
\end{equation}
in agreement with the large $N$ result \eqref{LargeNDimPhase1}. We can expand the classical action around this vev $s = s^* + t$, to obtain an action for the fluctuations 
\begin{equation}
\begin{split}
&S = - \frac{ 3 d^2 (d-2)^2}{32 g_{2} } \int d^d x \sqrt{g(x)} + \int d^d x \sqrt{g(x)} \bigg( \frac{(\partial_{\mu} t)^2}{2} + \frac{d (d-2)}{4} t^2 \\ 
& - \left( \bar{\Psi}_i  \left( \gamma\cdot \nabla + g_{1} s^* \right) \Psi^i + g_1 t \bar{\Psi}_i \Psi^i\right) + \frac{g_{2}}{6} s^* t^3  +  \frac{g_{2}}{24} t^4   \bigg).
\end{split}
\end{equation} 
Note that the fermion becomes massive now, with a mass given by $ g_{1} s^* $. According to our discussion in \ref{Sec:FreeMassiveFermion}, it leads to two possible boundary spinors with dimensions given by 
\begin{equation}
\hat{\Delta} = \frac{d-1}{2} \mp |g_1^* s^*|.
\end{equation} 
It is easy to see from \eqref{GNYVev} that $|g_1^* s^*| > 1$, so only the plus sign above is consistent with the boundary unitarity bound which requires boundary fermions to have dimensions greater than or equal to $(d - 2)/2$. At large $N$, it gives a $\hat{\Delta}$ which is consistent with \eqref{LargeNDimPhase1}. The fermion satisfies the boundary condition $\gamma_0 \Psi(z \rightarrow 0, \textbf{x}) = - \textrm{sgn} (g_1^* s^*)  \Psi(z \rightarrow 0, \textbf{x})$, which is also in agreement with the large $N$ result \eqref{LargeNFermTwoPP1}. The bulk mass of the scalar also gets shifted by the vev, so the dimension of the leading boundary scalar is now given by 
\begin{equation}
\hat{\Delta}_s (\hat{\Delta}_s - (d-1)) = \frac{d(d-2)}{2} \underset{d \rightarrow 4}{\rightarrow} 4,-1.
\end{equation}
We pick the unitary value $\hat{\Delta}_s  = 4$. In the large $N$ theory, this matches with the dimension of the leading operator that appears in boundary operator expansion of $\sigma$. This operator is proportional to the displacement operator and its dimension was referred to as $\hat{\Delta}_{(0)}$ in Table \ref{TableBoundaryPhasesDimensions}. The free energy in this phase is given by  
\begin{equation}
F = S_{\textrm{clas.}} + F_{t} + N F_{\Psi} =  - \frac{ 3 d^2 (d-2)^2 \textrm{Vol}(H^d)}{32 g_{2,0} } + F_{t} + N F_{\Psi}.
\end{equation}
We emphasize that we are using bare coupling here. This is because in $d = 4 - \epsilon$, the bare coupling gets renormalized as follows (see for instance \cite{Fei:2016sgs})
\begin{equation}
\frac{1}{g_{2,0}} = \mu^{- \epsilon} \left( \frac{1}{g_2} - \frac{3}{16 \pi^2 \epsilon} - \frac{ N g_1^2}{2 \pi^2 \epsilon g_2} + \frac{3 N g_1^4}{ \pi^2 \epsilon g_2^2}  . \right)
\end{equation}
The $1/\epsilon$ pole above has to be canceled by the other terms in the free energy. The field $t$ is a scalar with leading boundary operator of dimension $4$ in its spectrum. The free energy on hyperbolic space of such a scalar in $d = 4 - \epsilon$ is \cite{Giombi:2020rmc}
\begin{equation}
F_{t} = F_{\textrm{scalar}} \left( \frac{d}{2} \right) - \frac{9 \textrm{Vol} (H^4)}{8 \pi^2 \epsilon}.
\end{equation}
The fermion $\Psi$ is a massive fermion with mass $g_1^* s^*$ and its free energy is given by
\begin{equation} 
\begin{split}
F_{\Psi} = F_{\textrm{free}} + \int_0^{g_1^* s^*} dm \frac{\partial F}{\partial m} &= F_{\textrm{free}} -\frac{N c_d \textrm{Vol} (H^d) \Gamma \left( 1- \frac{d}{2} \right) }{(4 \pi)^{\frac{d}{2}}} \int_0^{g_1 s^*} dm \frac{\Gamma \left( \frac{d}{2} + m \right) }{ \Gamma \left( 1 - \frac{d}{2} + m \right)} \\
&= F_{\textrm{free}} - \frac{3 N g_1^2}{\pi^2 \epsilon g_2} + \frac{18 N g_1^4}{\pi^2 \epsilon g_2^2}. 
\end{split}
\end{equation}
Adding all the pieces together, we see that the $1/\epsilon$ pieces cancel and we get a finite free energy as a function of coupling 
\begin{equation}
\begin{split}
F &= N F_{\textrm{free}} + F_{\textrm{scalar}} \left( \frac{d}{2} \right) - \frac{6  \textrm{Vol} (H^4)}{g_2} \\
&= N F_{\textrm{free}} + F_{\textrm{scalar}} \left( \frac{d}{2} \right) - \frac{9 (2 N + 3)  \textrm{Vol} (H^4)}{4 \pi^2 \left( -2 N + 3 + \sqrt{4 N ^2 + 132 N + 9} \right) \epsilon } .
\end{split}
\end{equation} 
This is consistent with the large $N$ result in \eqref{Phase1Four}. 

In the phase where $s$ does not get a vev in the classical theory, we can choose to start from either Dirichlet or Neumann boundary condition on $s$ which corresponds to $B_2$ and $B_2'$ phases respectively. The one-point function of $s$ to leading order in the coupling may be calculated as 
\begin{equation} \label{IntegralSigmaOnePoint}
\langle s (x) \rangle = g_1 \int d^d x_1 \langle s (x) s (x_1) \rangle \langle \bar{\Psi}_i \Psi^i \rangle  (x_1). 
\end{equation}
The two-point function of $s$ in this phase is the usual one for a free scalar in hyperbolic space 
\begin{equation}
\langle s (x_1) s (x_2) \rangle^{N} = \frac{\Gamma\left( \frac{d}{2} - 1 \right)}{4 \pi ^{\frac{d}{2}}} \left( \frac{1}{\zeta^{\frac{d}{2} - 1}} + \frac{1}{(4 + \zeta)^{\frac{d}{2} - 1}} \right)
\end{equation}
when we impose Neumann boundary condition on $s$. When we impose Dirichlet, the only change is that there is a $-$ sign between the two terms. In both cases we need to do an integral of the following form in $d = 4 - \epsilon$
\begin{equation}
\begin{split}
\langle s (x) \rangle^N &= \mp \frac{g_1 N c_d \Gamma\left(\frac{d}{2} \right) \Gamma\left(\frac{d}{2} -1 \right)} { 2^{d + 2} \pi^{d} } \int d^{d - 1} \textbf{x}_1 d z_1  \bigg[ \frac{z^{\frac{d}{2} - 1} z_1^{-\frac{d}{2} - 1}}{((z_1 - z)^2 + (\textbf{x}_1 - \textbf{x})^2)^{\frac{d}{2} - 1}}  \\
& + \frac{z^{\frac{d}{2} - 1} z_1^{-\frac{d}{2} - 1}}{((z_1 + z)^2 + (\textbf{x}_1 - \textbf{x})^2)^{\frac{d}{2} - 1}} \bigg] \\
& = \pm \frac{g_1 N c_d \Gamma\left(\frac{d}{2} \right) } { 2^{d + 1} \pi^{\frac{d}{2}} } \int_0^{\infty} dz_1 z^{\frac{d}{2} - 1} z_1^{-\frac{d}{2} - 1} \left( |z - z_1| + (z + z_1) \right). 
\end{split}
\end{equation}
Again, for the Dirichlet case, there is a minus sign between the two terms. It is then easy to see that the integral of the second term is a pure divergence. It has a divergence at the boundary, $z_1 \rightarrow 0$, which must be cancelled by a boundary counterterm. But there is no finite part, so we can just set this integral to $0$. This is also expected, because this means that the one-point function is the same for Neumann and Dirichlet cases, and in the large $N$ theory, there is no distinction between Neumann and Dirichlet at the level of the one-point function of $\sigma$. The integral of the first term then gives  
\begin{equation}
\langle s(x) \rangle = \pm \frac{g_1 N}{8 \pi^2} \implies \sigma^* =   g_1^* \langle s (x) \rangle = \pm \frac{\epsilon N}{2 N + 3}
\end{equation}
which is in agreement with the large $N$ result \eqref{LargeNDimPhase2}. The free energy also depends on whether we choose Neumann or Dirichlet boundary conditions on the scalar. For Neumann, i.e. in the $B_2'$ phase, it is given by 

\begin{equation}
\begin{split}
F^N & = N F_{\textrm{Free}} + F_{\textrm{scalar}} \left( \frac{d}{2} - 1 \right) + \frac{g_2 \textrm{Vol}(H^d)}{8} \langle s^2 \rangle^2 - \frac{g_1^2}{2} \int d^d x_1 d^d x_2 \langle s \bar{\Psi}_i \Psi^i (x_1) s \bar{\Psi}_j \Psi^j (x_2) \rangle \\
& =  N F_{\textrm{Free}} + F_{\textrm{scalar}} + \frac{g_2 \textrm{Vol}(H^4) }{8 (4 \pi)^{4}} - \frac{g_1^2 N^2 \textrm{Vol}(H^4)  }{128 \pi^{6}}  \int d^d x \sqrt{g_x} \left( \frac{1}{\zeta^{\frac{d}{2} -1}} + \frac{1}{ (4 + \zeta)^{\frac{d}{2} -1}}\right) \\
& - \frac{g_1^2 \textrm{Vol}(H^4) N }{8 \pi^{6}} \int d^d x \sqrt{g_x} \left( \frac{1}{\zeta^{d -1}} - \frac{1}{(4 + \zeta)^{d - 1}} \right) \left( \frac{1}{\zeta^{\frac{d}{2} -1}} + \frac{1}{(4 + \zeta)^{\frac{d}{2} - 1}}  \right)
\end{split}
\end{equation} 
where we already fixed one of the points at the center of hyperbolic space, and the integral over that point resulted in a factor of the volume of the hyperbolic space. The integral in the first line is the same as what we did in \eqref{IntegralSigmaOnePoint}, so we can just use the same result. To do the integral in the second line, it is convenient to use ball coordinates on the hyperbolic space. In these coordinates, the metric is given by 
\begin{equation}
ds^2 = \frac{4}{(1- u^2)^2} \left( du^2 + u^2 d \Omega^2_{d-1} \right).
\end{equation}
We then fix one of the points at the center of the ball. The chordal distance in terms of $u$ variables between the center to an arbitrary point is given by 
\begin{equation}
\zeta = \frac{4 u^2}{1 - u^2}.
\end{equation}
The integral in $d = 4$ then gives 
\begin{equation}
\begin{split}
J =  - \frac{g_1^2 \textrm{Vol}(H^4) N }{64 \pi^{4}} \int_0^1 d u \ u^{d-1} (1- u^2)^{\frac{d}{2} - 2} (u^{2 - 2 d} - 1) (u^{2 - d} + 1) = \frac{7 g_1^2 N \textrm{Vol}(H^4)}{512 \pi^4}.
\end{split}
\end{equation}
Putting all the pieces together, the free energy to leading order in $\epsilon$ is given by
\begin{equation}
\begin{split}
F^N &= N F_{\textrm{Free}} \ + \ F_{\textrm{scalar}} \left( \frac{d}{2} - 1 \right) \ + \ \frac{N^2 \textrm{Vol}(H^4) \epsilon}{8 \pi^2 (2 N + 3)} + \frac{7 N \textrm{Vol}(H^4) \epsilon}{(2 N + 3) 64 \pi^2}  \\
& + \  \frac{ \textrm{Vol}(H^4) \left( - 2 N + 3  + \sqrt{4 N ^2 + 132 N + 9} \right)  \epsilon }{48 (2 N + 3) (4 \pi)^{2}}. 
\end{split}
\end{equation}
In the $B_2$ phase, when we choose Dirichlet boundary condition on $s$, we get the following result for the free energy

 \begin{equation}
\begin{split}
F^D & = N F_{\textrm{Free}} + F_{\textrm{scalar}} \left( \frac{d}{2} \right) + \frac{g_2 \textrm{Vol}(H^4) }{8 (4 \pi)^{4}} - \frac{g_1^2 N^2 \textrm{Vol}(H^4)  }{128 \pi^{6}}  \int d^d x \sqrt{g_x} \left( \frac{1}{\zeta^{\frac{d}{2} -1}} - \frac{1}{ (4 + \zeta)^{\frac{d}{2} -1}}\right) \\
& - \frac{g_1^2 \textrm{Vol}(H^4) N }{8 \pi^{6}} \int d^d x \sqrt{g_x} \left( \frac{1}{\zeta^{d -1}} - \frac{1}{(4 + \zeta)^{d - 1}} \right) \left( \frac{1}{\zeta^{\frac{d}{2} -1}} - \frac{1}{(4 + \zeta)^{\frac{d}{2} - 1}}  \right) \\
&= N F_{\textrm{Free}} \ + \ F_{\textrm{scalar}} \left( \frac{d}{2} \right) \ +  \ \frac{N^2 \textrm{Vol}(H^4) \epsilon}{8 \pi^2 (2 N + 3)} + \frac{3 N \textrm{Vol}(H^4) \epsilon}{(2 N + 3) 64 \pi^2}   \\
& + \frac{ \textrm{Vol}(H^4) \left( - 2 N + 3  + \sqrt{4 N ^2 + 132 N + 9} \right)  \epsilon }{48 (2 N + 3) (4 \pi)^{2}}.
\end{split}
\end{equation}
In both $B_2$ and $B_2'$ phases, the order $N$ piece in the free energy at large $N$ agrees with the result of the large $N$ calculation \eqref{Phase2Four}. However, they differ at order $1$ with the difference given by
\begin{equation}
\begin{split}
F^N - F^D &= F_{\textrm{scalar}} \left( \frac{d}{2} - 1 \right) - F_{\textrm{scalar}} \left( \frac{d}{2} \right) + \frac{ N \textrm{Vol}(H^4) \epsilon}{(2 N + 3) 16 \pi^2} \\
&= - \frac{1}{\sin \left( \frac{\pi (d - 1)}{2} \right) \Gamma(d)} \int_0^{\frac{1}{2}} d u u \sin \pi u \Gamma\left( \frac{d - 1}{2} + u \right) \Gamma\left( \frac{d - 1}{2} - u \right) + \frac{ N \epsilon}{(2 N + 3) 12}
\end{split}
\end{equation}
where we used the general formula \eqref{DoubleTraceScalar} to calculate the difference in the free energy of a free scalar with dimension $d/2 - 1$ and $d/2$. This difference is also in agreement with the large $N$ result in \eqref{LargeNNeumannDirichlet}.

\section{Using equations of motion in the bulk} \label{Sec:EOM}
In this section, we turn to bulk correlation functions. Since we have a Lagrangian description of our models, the bulk fields satisfy equations of motion, which in turn implies that the correlation functions involving bulk fields must satisfy a differential equation. This differential equation can be solved in some situations to yield the correlation function. Such an approach was originally used to calculate anomalous dimensions in a CFT in \cite{Rychkov:2015naa} and was later extended to calculate two-point functions in a BCFT \cite{Giombi:2020rmc} and in a CFT on real projective space \cite{Giombi:2020xah}. This is an alternative approach to calculating Feynman diagrams in half-space or in AdS. We start with the correlation functions involving the scalar $s$ in GNY model, where the calculation is very similar to \cite{Giombi:2020rmc, Giombi:2020xah}. We then move on to the correlation functions involving the fermion and fix the fermion two-point function in both the GN model and the GNY model, to leading order in $\epsilon$.  

\subsection{Scalar}
Let us look at the $s$ correlator in the GNY model \eqref{ActionGNYAdS} in $d = 4 - \epsilon$ in the phase where $s$ does not get a vev (classically). As a warm up, we start with the bulk-boundary propagator on $H^d$ which must take the following form 
\begin{equation}
\langle s(x_1) \hat{s}(\textbf{x}_2) \rangle = B_{s \hat{s}} \left(  \frac{z_1}{z_1^2 + \textbf{x}_{12}^2} \right)^{\hat{\Delta}_s}.
\end{equation}
Applying the $H^d$ equation of motion at $x_1$ gives 
\begin{equation} \label{EOMScalarLHS}
\left( \nabla_{x_1}^2 + \frac{d (d - 2)}{4} \right) \langle s(x_1) \hat{s}(\textbf{x}_2) \rangle =  \frac{ ( 2 \hat{\Delta}_s - d ) ( 2 \hat{\Delta}_s - d + 2 ) }{4} B_{s \hat{s}} \left(  \frac{z_1}{z_1^2 + \textbf{x}_{12}^2} \right)^{\hat{\Delta}_s}.
\end{equation} 
In the free theory, the right hand side above must be set to $0$ which gives the usual boundary dimensions for Neumann and Dirichlet boundary conditions. In the interacting GNY model, the equation of motion, to leading order in $\epsilon$ gives 
\begin{equation} \label{EOMScalarRHS}
\begin{split}
\left( \nabla_{x_1}^2 + \frac{d (d - 2)}{4} \right) \langle s(x_1) \hat{s}(\textbf{x}_2) \rangle &= \frac{g_2}{2} \langle s^2 (x_1) \rangle  \langle  s(x_1) \hat{s}(\textbf{x}_2)  \rangle - g_1^2 \langle \bar{\Psi}_{i} \Psi^{i} \rangle^2  \int d^d x \sqrt{g_x} \langle s  (x) \hat{s} (\textbf{x}_2) \rangle \\
&- g_1^2 \int d^d x \sqrt{g_x} \langle \bar{\Psi}_{i \alpha} (x_1) \Psi^{j \beta} (x) \rangle \langle \Psi^{i \alpha} (x_1) \bar{\Psi}_{j \beta} (x) \rangle \langle s  (x) \hat{s} (\textbf{x}_2) \rangle. 
\end{split}
\end{equation}
Comparing \eqref{EOMScalarLHS} and \eqref{EOMScalarRHS} should give us the anomalous dimension of the leading boundary scalar $\hat{s}$ to leading order in $\epsilon$. So let us try to evaluate the right hand side of \eqref{EOMScalarRHS}. The first term is straightforward. As for the second term, it is easy to see that the integral should be set to $0$ (it is pure power law divergence at $z\rightarrow 0$ but this can be absorbed in a boundary counterterm ). The last term is non-trivial and the integral involved is as follows 
\begin{equation}
\begin{split}
I &= - \frac{g_1^2 \Gamma \left( \frac{d}{2} \right)^2 N c_d B_{s \hat{s}} }{4 \pi^{d} } \int d^d x \sqrt{g_x} \left( \frac{1}{\zeta_{x_1 x}^{d - 1}} - \frac{1}{ (4 + \zeta_{x_1 x} )^{d - 1} } \right) \left(  \frac{z}{z^2 + (\textbf{x} - \textbf{x}_2)^2} \right)^{\hat{\Delta}_s} \\
&=  - \frac{g_1^2 \Gamma \left( \frac{d}{2} \right)^2 N c_d z_1^{d-1} B_{s \hat{s}} }{4 \pi^{d} } \int d z d^{d-1} \textbf{x} \bigg[  \frac{z^{\hat{\Delta}_s - 1}}{ ((z- z_1)^2 + (\textbf{x} - \textbf{x}_1)^2)^{d - 1} (z^2 + (\textbf{x} - \textbf{x}_2)^2)^{\hat{\Delta}_s}} \\
& -  \frac{z^{\hat{\Delta}_s -1}}{ ((z + z_1)^2 + (\textbf{x} - \textbf{x}_1)^2)^{d - 1} (z^2 + (\textbf{x} - \textbf{x}_2)^2)^{\hat{\Delta}_s}} \bigg] 
\end{split}
\end{equation}
The integral over $\textbf{x}$ can be performed using Feynman parameters, and then the integral over the Feynman parameter can be performed leaving us with the following integral over $z$

\begin{equation}
\begin{split}
I &= - \frac{g_1^2 \Gamma \left( \frac{d}{2} \right)^2 N c_d \Gamma \left( \frac{d-1}{2}\right)   B_{s \hat{s}} }{4 \pi^{\frac{d + 1}{2}} \Gamma (d-1) } \frac{z_1^{d-1}}{(z_1^2 + \textbf{x}_{12}^2)^{\hat{\Delta}_s}} \int d z \left( \frac{z^{\hat{\Delta}_s - 1}}{((z - z_1)^2)^{\frac{d - 1}{2}}} - \frac{z^{\hat{\Delta}_s - 1}}{((z + z_1)^2)^{\frac{d - 1}{2}}} \right) \\
&= - \frac{g_1^2 \Gamma \left( \frac{d}{2} \right)^2 N c_d \Gamma \left( \frac{d-1}{2}\right)   B_{s \hat{s}} }{4 \pi^{\frac{d + 1}{2}} \Gamma (d-1) } \left( \frac{z_1}{z_1^2 + \textbf{x}_{12}^2} \right)^{\hat{\Delta}_s} \bigg[ -   \frac{\Gamma \left( d - 1 - \hat{\Delta}_s \right) \Gamma \left( \hat{\Delta}_s \right)}{ \Gamma (d-1)}  \\
&+ \Gamma (2-d) \left(\frac{\Gamma (d - \hat{\Delta}_s -1)}{\Gamma (1-\hat{\Delta}_s )}+\frac{\Gamma (\hat{\Delta}_s )}{\Gamma (-d + \hat{\Delta}_s +2)}\right) \bigg]
\end{split}
\end{equation}
where the first integral only converges for $d < 2, \hat{\Delta}_s > 0 $ and $d - \hat{\Delta}_s > 1$, but the final answer can be analytically continued in $d$ and $\hat{\Delta}_s$. The two terms in the second line add up to $0$ for both Neumann and Dirichlet cases, i.e. for both  $\hat{\Delta}_s = d/2 -1$ and $d/2$, so the resulting integral for these two cases becomes 
\begin{equation}
I =  \frac{g_1^2 \Gamma \left( \frac{d}{2} \right)^2 N c_d \Gamma \left( \frac{d-1}{2}\right) \Gamma \left( d - 1 - \hat{\Delta}_s\right) \Gamma \left( \hat{\Delta}_s \right) }{4 \pi^{\frac{d + 1}{2}} \Gamma (d-1)^2 } B_{s \hat{s}}  \left( \frac{z_1}{z_1^2 + \textbf{x}_{12}^2} \right)^{\hat{\Delta}_s}.
\end{equation}
Using this, and the fact that in the free theory in $d = 4$, $\langle s^2 (x) \rangle = \pm 1/(4 \pi)^2$ we can calculate the dimensions of the leading boundary scalar in GNY model 
\begin{equation} \label{AnomalousOneBoxScalN}
\begin{split}
&\hat{\gamma}^N_s = -\frac{g_2}{2 (4 \pi)^2}  -\frac{g_1^2  N c_d }{8 \pi^{2}} = - \frac{\sqrt{4 N^2 + 132 N + 9} + 10 N + 3}{12 (2 N + 3)} \epsilon \\
&\hat{\Delta}^N_s = \frac{d}{2} - 1 + \hat{\gamma}^N_s  = 1  - \frac{\sqrt{4 N^2 + 132 N + 9} + 22 N + 21}{12 (2 N + 3)} \epsilon
\end{split}
\end{equation}
in $B_2'$ phase and 
\begin{equation} \label{AnomalousOneBoxScalD}
\begin{split}
&\hat{\gamma}^D_s = -\frac{g_2}{2 (4 \pi)^2}  +\frac{g_1^2  N c_d }{8 \pi^{2} } = - \frac{\sqrt{4 N^2 + 132 N + 9} - 14 N + 3}{12 (2 N + 3)} \epsilon \\
&\hat{\Delta}^D_s = \frac{d}{2}  + \hat{\gamma}^D_s  = 2  - \frac{\sqrt{4 N^2 + 132 N + 9} - 2 N + 21}{12 (2 N + 3)} \epsilon
\end{split}
\end{equation}
in the $B_2$ phase. At leading order in large $N$, they are equal $1 - \epsilon$ and $2$ respectively, in agreement with the large $N$ values of $d - 3$ and $2$. 

Next, we look at the two-point function of $s$, in which case, we can apply the equation of motion at both points. In this case, to leading order in the perturbation theory, we get the following differential equation for the two-point function
\begin{equation} 
\begin{split}
&\left( \nabla_{x_2}^2 + \frac{d (d - 2)}{4} \right) \left( \nabla_{x_1}^2 + \frac{d (d - 2)}{4} \right) \langle s(x_1) s(x_2) \rangle =   \\
& g_1^2 \langle \bar{\Psi}_{i} \Psi^{i} \rangle^2  + g_1^2 \langle \bar{\Psi}_{i a} (x_1) \Psi^{j b} (x_2) \rangle \langle \Psi^{i a} (x_1) \bar{\Psi}_{j b} (x_2) \rangle. 
\end{split}
\end{equation}
Writing the propagator as a function of $\zeta$ , $G_{s} (\zeta)$, we get following differential equation for the propagator, keeping only terms to order $\epsilon$ on the RHS
 \begin{equation} \label{TwoBoxGNY}
\begin{split}
&\bigg[\zeta (4 + \zeta) \left( \zeta (4 + \zeta) \partial_{\zeta}^4 + (d + 2)(4 + 2 \zeta) \partial_{\zeta}^3 \right) + \frac{\left( 8 d(d + 2) + (4 + 3 d (d + 2)) \zeta (4 + \zeta) \right)}{2} \partial_{\zeta}^2 + \\
& \frac{d^3 (4 + 2 \zeta)}{4} \partial_{\xi} + \frac{d^2 (d - 2)^2}{16} \bigg] G_{s} (\zeta) = D^{(4)} G_{s} (\zeta) = \frac{ g_1^2 N c_d  \Gamma \left( \frac{d}{2} \right)^2 }{4 \pi^d} \left[\frac{N c_d}{4^{d - 1}} +  \frac{1}{\zeta^{d - 1}} - \frac{1}{ (4 + \zeta)^{d - 1} }  \right].
\end{split}
\end{equation} 
Recall from \eqref{BlockDecompTwoPS} the conformal block decomposition for the two-point function. It turns out that the boundary channel block is an eigenfunction of the equation of motion operator 
\begin{equation}
D^{(4)} f_{\textrm{bdry}} ({\hat{\Delta}}_l;\zeta) = \frac{\left(d - 2 {\hat{\Delta}}_l \right)^2  \left(d -2 - 2 {\hat{\Delta}}_l \right)^2}{16} f_{\textrm{bdry}} ({\hat{\Delta}}_l;\zeta).
\end{equation}
This allows us to plug in the block decomposition into \eqref{TwoBoxGNY} and extract information about the bulk-boundary OPE coefficients at order $\epsilon$. For instance, the one-point function of $s$ is fixed to be 
\begin{equation}
A \frac{(d - 2)^2 d^2}{16} a_s^2 =  \frac{ g_1^2 N^2 c_d^2  \Gamma \left( \frac{d}{2} \right)^2 }{(4 \pi)^d}.
\end{equation}
The boundary expansion coefficients of all the subleading boundary operators obey the following constraint
\begin{equation}
A\sum_l \frac{\left(d - 2 {\hat{\Delta}}_l \right)^2  \left(d -2 - 2 {\hat{\Delta}}_l \right)^2}{16} \mu_l^2 f_{\textrm{bdry}} ({\hat{\Delta}}_l;\zeta) = \frac{ g_1^2 N c_d  \Gamma \left( \frac{d}{2} \right)^2 }{4 \pi^d} \left( \frac{1}{\zeta^{3}} - \frac{1}{ (4 + \zeta)^{3} }  \right)
\end{equation}
where the sum does not include the leading boundary operator of dimension $d/2-1$ or $d/2$. In the usual normalization, $A = 1/4 \pi^2$ in four dimensions. Expanding both sides in powers of $\zeta$ tells us that the boundary spectrum contains a tower of operators of dimension $4 + 2 n$ with the OPE coefficients 
\begin{equation}
\mu_{4 + 2 n}^2 = \frac{N  \Gamma(2 n + 2) \sqrt{\pi} \epsilon}{ (2 N + 3) 2^{4 n + 5} \Gamma \left( 2 n + \frac{5}{2} \right)}. \ \  
\end{equation}
This is consistent with what we found in the large $N$ expansion \eqref{BulkBounSigB2}. The $n = 0$ operator in the tower is proportional to the displacement operator in $B_2$ and $B_2'$ phases in this description.  

We can also directly solve the equation \eqref{TwoBoxGNY} perturbatively in $\epsilon$ by expanding the differential operator and the correlator in powers of $\epsilon$
\begin{equation}
\begin{split}
D^{(4)} &= D^{(4)}_0 + \epsilon D^{(4)}_1 + O(\epsilon^2) \\
G_{s} (\zeta) &= G_{0} (\zeta) + \epsilon G_{1} (\zeta) + O(\epsilon^2). 
\end{split}
\end{equation}
Let us now work in a more convenient normalization where the free theory correlator is \footnote{Note that when we change the normalization of fields, the coupling constant also needs to change accordingly. So in this normalization, coupling constant $g_1$ changes to $g_1 (2 \pi)$ to leading order in $\epsilon$.}
\begin{equation}
G^{N/D}_{0} (\zeta) = \frac{1}{\zeta} \pm \frac{1}{4 + \zeta}. 
\end{equation}
We then get the following differential equation for $G_{1} (\zeta)$
\begin{equation}
D^{(4)}_0 G_{1} (\zeta) = \frac{3 2 N}{2 N + 3} \left( \frac{N}{16} + \frac{1}{\zeta^3} - \frac{1}{(4 + \zeta)^3}  \right) - D^{(4)}_1 G_{0} (\zeta). 
\end{equation}
The equation can be solved to give 
\begin{equation} 
\begin{split}
G^N_{1} (\zeta) &= \frac{c_1}{\zeta} + \frac{c_2}{4 + \zeta} + \frac{c_3 \log \left( \zeta /4\right) }{4 + \zeta} + \frac{c_4 \log  (1 + \zeta/4)}{\zeta} + \frac{N^2}{2 (2 N + 3)} \\
&+ \frac{3}{2 (2 N + 3)} \left( \frac{\log \zeta}{\zeta} +  \frac{\log (4 + \zeta)}{4 + \zeta} \right)  + \ \frac{2 N}{(2 N + 3)} \frac{\log (1 + \zeta/4)}{4 + \zeta}  \\
G^D_{1} (\zeta) &= \frac{c_1}{\zeta} + \frac{c_2}{4 + \zeta} + \frac{c_3 \log \left( \zeta /4\right)}{4 + \zeta} + \frac{c_4 \log  (1 + \zeta/4)}{\zeta} + \frac{N^2}{2 (2 N + 3)} \\
&+ \frac{3}{2 (2 N + 3)} \left( \frac{\log \zeta}{\zeta} -  \frac{\log (4 + \zeta)}{4 + \zeta} \right). 
\end{split}
\end{equation}
We have four undetermined coefficients. One of these is fixed by fixing the normalization of the field $s$. We are working in a normalization such that the correlator falls off as $1/ \zeta$ as $\zeta \rightarrow 0$, which sets $c_1 = 0$ for both Neumann and Dirichlet cases. For further analysis, we have to consider Dirichlet and Neumann cases separately. For the Dirichlet case, the leading boundary operator has dimension $2$, so the large $\zeta$ expansion of the two-point function should not have any $1/\zeta$ or $(\log \zeta) /\zeta$ terms. This implies that $c_2 = 0$ and $c_4 = - c_3$. So we are left with one undetermined coefficient. This can be fixed by looking at the bulk OPE limit ($\zeta \rightarrow 0$), where the correlator should behave like 
\begin{equation} \label{BulkChannelS}
\begin{split}
G_{s} (\zeta) &= \zeta^{-\Delta_s} + \lambda_{s^2} \zeta^{\frac{1}{2} ( \Delta_{s^2} - 2 \Delta_s )} + \textrm{higher orders in } \zeta \\
&= \zeta^{-1} + \lambda_{s^2}^{(0)} + \left( \left(\frac{1}{2} - \gamma_s^{(1)} \right) \frac{\log \zeta}{\zeta} + \lambda_{s^2}^{(1)} + \left(\frac{\gamma^{(1)}_{s^2}}{2} - \gamma_s^{(1)} \right) \lambda_{s^2}^0  \log \zeta \right) \epsilon.
\end{split}
\end{equation}
Free theory result fixed $\lambda_{s^2}^0 =  - 1/4$. Then we get the following result by comparing $\log \zeta$ terms 
\begin{equation}
\frac{\gamma^{(1)}_{s^2}}{2} - \gamma_s^{(1)} = -c_3
\end{equation}
Using the following bulk data from \cite{Fei:2016sgs}, we can calculate $c_3$
\begin{equation}
\Delta_{s} = 1 - \frac{3}{2 (2 N + 3 )} \epsilon, \hspace{1 cm} \Delta_{s^2} = 2 + \frac{\sqrt{4 N^2 + 132 + 9} - 2 N -15}{6 (2 N + 3)} \epsilon
\end{equation}
and we get 
\begin{equation}
c_3 = \frac{-\sqrt{4 N^2 + 132 N + 9} + 2 N - 3}{12 (2 N + 3)}
\end{equation}
This fixes the $s$ two-point function in the $B_2$ phase to be 
\begin{equation}
G^D_{1} (\zeta) = c_3 \left( \frac{\log \left( \zeta /4\right)}{4 + \zeta} - \frac{\log  (1 + \zeta/4)}{\zeta} \right) + \frac{N^2}{2 (2 N + 3)} + \frac{3}{2 (2 N + 3)} \left( \frac{\log \zeta}{\zeta} -  \frac{\log (4 + \zeta)}{4 + \zeta} \right). 
\end{equation}
At large $N$, this agrees with what we found in \eqref{SigmaCorrEps}. This determines all the BCFT data to order $\epsilon$. In particular, the dimension and boundary expansion coefficient of the leading boundary scalar is 
\begin{equation}
\hat{\Delta}_s^D = 2  - \frac{\sqrt{4 N^2 + 132 N + 9} - 2 N + 21}{12 (2 N + 3)} \epsilon, \hspace{0.5cm} \mu_2^2 = \frac{1}{4} + \left( \frac{c_3 (2 \log 2 - 1)}{4} - \frac{3}{8 (2 N + 3)} \right) \epsilon
\end{equation}
which agrees with what we found above \eqref{AnomalousOneBoxScalD} and is also consistent with the large $N$ result \eqref{BulkBounSigB2}.

Next, we consider the $B_2'$ phase where the leading boundary operator has dimension $1$ and the next subleading operator has dimension $4$ in the free theory. This implies that $1/\zeta^2$ and $(\log \zeta)/\zeta^2$ terms must be descendants of the leading operator (similarly for $1/\zeta^3$). This puts constraints on the coefficients  
\begin{equation}
c_2 = \frac{4 N}{3 + 2 N}, \hspace{1cm} c_4 = c_3 + \frac{2 N}{3 + 2 N}. 
\end{equation}
We then compare with the bulk channel expansion \eqref{BulkChannelS}. The free theory result implies $\lambda_{s^2}^0 =  - 1/4$ and comparing the coefficient of $\log \zeta$ gives 
\begin{equation}
\frac{\gamma^{(1)}_{s^2}}{2} - \gamma_s^{(1)} = c_3 \implies c_3 = \frac{\sqrt{4 N^2 + 132 N + 9} - 2 N + 3}{12 (2 N + 3)}.
\end{equation}
So the full $O(\epsilon)$ correlator in the $B_2'$ phase is the following 
\begin{equation} 
\begin{split}
G^N_{1} (\zeta) &=  \frac{4 N}{(3 + 2 N)} \frac{1}{4 + \zeta} + c_3 \left( \frac{ \log \left( \zeta /4\right) }{4 + \zeta} + \frac{\log  (1 + \zeta/4)}{\zeta} \right) + \frac{N^2}{2 (2 N + 3)} \\
&+ \frac{3}{2 (2 N + 3)} \left( \frac{\log \zeta}{\zeta} +  \frac{\log (4 + \zeta)}{4 + \zeta} \right)  + \ \frac{2 N}{(2 N + 3)} \left( \frac{1}{\zeta} + \frac{1}{4 + \zeta} \right) \log (1 + \zeta/4).
\end{split}
\end{equation}
This also agrees with the large $N$ result \eqref{SigmaCorrEps}. The dimension and boundary expansion coefficient of the leading boundary scalar in this case is 
\begin{equation} 
\hat{\Delta}_s^N = 1  - \frac{\sqrt{4 N^2 + 132 N + 9} + 22 N + 21}{12 (2 N + 3)} \epsilon, \hspace{0.5 cm} \mu_1^2 = \frac{1}{2} + \frac{\epsilon N}{2 N + 3} 
\end{equation}
again in agreement with \eqref{AnomalousOneBoxScalN} and with the large $N$ result \eqref{BulkBounSigB2'}.

\subsection{Fermion}
In this subsection, we apply the same logic to fermion correlators. As we wrote in \eqref{FermionBulkBoundaryGen}, the bulk-boundary propagator of a fermion can be written as 
\begin{equation}
\langle \Psi(x_1) \hat{\bar{\Psi}} (\textbf{x}_2) \rangle = B_{\Psi \hat{\Psi}}  \frac{ \gamma_{a} x_{12}^a \left( 1 \mp \gamma_0 \right) z_1^{\hat{\Delta}}}{ (z_1^2 + \textbf{x}_{12}^2)^{\hat{\Delta} + 1/2 }}
\end{equation} 
where the fermion satisfies the boundary condition $\gamma_0 \Psi (z_1 \rightarrow 0, \textbf{x}_1) = \pm  \Psi (z_1 \rightarrow 0, \textbf{x}_1)$. Acting with the Dirac operator on the right hand side above gives 
\begin{equation} \label{EOMOneBoxFermionLHS}
\gamma \cdot \nabla_1 \langle \Psi(x_1) \hat{\bar{\Psi}} (\textbf{x}_2) \rangle = \left( z_1 \gamma^a \partial_{1a} - \frac{(d - 1)}{2} \gamma_0 \right)\langle \Psi(x_1) \hat{\bar{\Psi}} (\textbf{x}_2) \rangle = \pm \left( \hat{\Delta} - \frac{d - 1}{2} \right) \langle \Psi(x_1) \hat{\bar{\Psi}} (\textbf{x}_2) \rangle
\end{equation}
For the free massive fermion, the equation of motion sets $\gamma \cdot \nabla \Psi = - m \Psi$ which gives the dimension of the leading boundary spinor $\hat{\Delta} = (d-1)/2 \mp m$. In the Gross-Neveu model, the equation of motion sets 
\begin{equation} \label{EOMOneBoxFermionRHSGN}
\gamma \cdot \nabla_1 \langle \Psi^i (x_1) \hat{\bar{\Psi}}_j (\textbf{x}_2) \rangle = - g \left(N - \frac{1}{2} \right) \langle \bar{\Psi} \Psi \rangle \langle \Psi^i(x_1) \hat{\bar{\Psi}}_j (\textbf{x}_2) \rangle
\end{equation}  
Since there is an explicit factor of $g$ on the right hand side above, we can plug in the one-point function and the correlator for the free theory, and on comparing it with \eqref{EOMOneBoxFermionLHS}, this should give us the anomalous dimension for the leading boundary spinor. Using the one-point function from \eqref{PsibarPsiOnePoint}, we get, in $d = 2 + \epsilon$,
\begin{equation}
\pm \left( \hat{\Delta} - \frac{d - 1}{2} \right) = \pm  g^* \left(N - \frac{1}{2} \right)  \frac{c_d \Gamma \left(\frac{d}{2} \right)}{(4 \pi)^{\frac{d}{2}}} \implies \hat{\Delta} =  \frac{1}{2} + \frac{4 N - 3}{4 (N - 1)} \epsilon
\end{equation}
in agreement with the result we got by direct calculation \eqref{BoundaryFAnomGN}.

We now turn to the bulk two-point function of the fermion. We start with the ansatz in \eqref{FermionTwoPointAnsatz} and act on it with the Dirac operator which gives \eqref{DiracOperatorAnsatz}. In the free theory, the equation of motion sets this derivative to zero away from the coincident limit, which gives us two first order differential equations
\begin{equation}
\begin{split}
\gamma \cdot \nabla_1  G_{\Psi} (x_1, x_2) = 0 \implies  \alpha'(\zeta) + \frac{(d - 1)}{2} \frac{\alpha (\zeta)}{\zeta + 4} = 0, \hspace{0.5cm} \beta'(\zeta) + \frac{(d - 1)}{2} \frac{\beta (\zeta)}{\zeta} =0.
\end{split}
\end{equation}
These equations can be solved to give 
\begin{equation}
\beta(\zeta) = \frac{c_1}{\zeta^{\frac{d-1}{2}}}, \hspace{0.5cm} \alpha(\zeta) = \frac{c_2}{(\zeta + 4)^{\frac{d-1}{2}}} .
\end{equation}
One of these constants can be fixed by fixing the overall normalization of the two-point function. For convenience, we now work with the convention such that as $\zeta \rightarrow 0$, the two-point function goes like $-(\gamma_a x_{12}^a)/ \zeta^{\frac{d}{2}}$, which sets $c_1 = - 1$. We then recall that the boundary condition requires 
\begin{equation} \label{FermionBoundaryCondition}
\gamma_0 G_{\Psi} (x_1, x_2) |_{z_1 \rightarrow 0} = \pm G_{\Psi} (x_1, x_2) |_{z_1 \rightarrow 0} \implies c_2 = \pm 1
\end{equation}

In the Gross-Neveu model, the equation of motion requires 
\begin{equation} \label{EOMGN}
\gamma \cdot \nabla_1  G_{\Psi} (x_1, x_2) = -g \left(N - \frac{1}{2} \right) \langle \bar{\Psi} \Psi \rangle G_{\Psi} (x_1, x_2).
\end{equation}
We can then solve this equation perturbatively in $d = 2 + \epsilon$ by expanding
\begin{equation}
\alpha (\zeta) = \alpha_0(\zeta) + \epsilon \alpha_1(\zeta), \hspace{0.25cm} \alpha_0(\zeta) = \pm \frac{1}{\sqrt{\zeta + 4}} ; \hspace{0.5cm} \beta (\zeta) = \beta_0(\zeta) + \epsilon \beta_1(\zeta),  \hspace{0.25cm} \beta_0(\zeta) = - \frac{1}{\sqrt{\zeta}}. 
\end{equation}
Plugging this into \eqref{EOMGN} and comparing the coefficients of $\gamma_a x_{12}^a$ and $\gamma_0 \gamma_a \bar{x}_{12}^a$ gives the following equations at order $\epsilon$
\begin{equation} 
\begin{split}
\alpha'_1(\zeta) +\frac{\alpha_1(\zeta)}{2 (\zeta + 4)} &= \mp \frac{\zeta + 2}{\zeta (\zeta + 4)^{3/2}} \mp \frac{1}{4 (N - 1) \zeta \sqrt{\zeta + 4}}; \\
\beta'_1(\zeta) +\frac{\beta_1(\zeta)}{2 \zeta } &= \frac{\zeta + 2 }{\zeta^{3/2} (\zeta + 4)} + \frac{1}{4 (N - 1) \sqrt{\zeta} (\zeta + 4)}.
\end{split}
\end{equation}
The solutions are 
\begin{equation}
\begin{split}
\beta_1(\zeta) &= \frac{d_1}{\sqrt{\zeta}} + \frac{1}{2 \sqrt{\zeta}} \log \left( \frac{\zeta (\zeta + 4)}{16} \right) + \frac{\log (\zeta/4 + 1)}{4 (N - 1) \sqrt{\zeta}}; \\
\alpha_1(\zeta) &=  \frac{d_2}{\sqrt{\zeta + 4}} \mp  \frac{1}{2 \sqrt{\zeta + 4}} \log \left( \frac{\zeta (\zeta + 4)}{16} \right) \mp  \frac{ \log (\zeta /4)}{4 (N - 1) \sqrt{\zeta + 4}}.
\end{split}
\end{equation}
Fixing the normalization sets $d_1 = 0$. And then requiring that the boundary condition \eqref{FermionBoundaryCondition} is satisfied as $z_1 \rightarrow 0$ fixes $d_2 = 0$. So the full correlator, to order $\epsilon$ in GN model is 
\begin{equation}
\begin{split} 
G_{\Psi} (x_1, x_2) &= \frac{\gamma_0 \gamma_{a} (\bar{x}_1 - x_2)^{a} }{ \sqrt{z_1 z_2}} \left( \mp \frac{1}{\zeta + 4} \pm  \frac{\epsilon}{2 (\zeta + 4)} \log \left( \frac{\zeta (\zeta + 4)}{16} \right) \pm \epsilon \frac{ \log (\zeta /4)}{4 (N - 1) (\zeta + 4)} \right) \\
& + \frac{\gamma_{a} (x_1 - x_2)^{a} }{\sqrt{z_1 z_2}} \left( - \frac{1}{\zeta} +  \frac{\epsilon}{2 \zeta}\log \left( \frac{\zeta (\zeta + 4)}{16} \right) + \epsilon \frac{\log (\zeta/4 + 1)}{4 (N - 1) \zeta}  \right).
\end{split}
\end{equation}
At large $N$, this agrees with the fermion two-point function we found at large $N$ in $B_1$ phase \eqref{LargeNFermTwoPP1}. As a check, looking at the coefficient of $\log \zeta$ in $\zeta \rightarrow \infty$ limit, we recover the dimension of the leading boundary fermion 
\begin{equation}
\hat{\Delta} = \frac{1}{2} + \frac{4 N - 3}{4 (N - 1)} \epsilon.
\end{equation}

In the GNY model, it is more convenient to apply the Dirac operator on both of the fermions in the two-point function. Acting on the ansatz \eqref{FermionTwoPointAnsatz} with two Dirac operators we get
\begin{equation}
\begin{split}
&\left( z_1 \gamma^a \partial_{1a} - \frac{(d - 1)}{2} \gamma_0 \right) G_{\Psi} (x_1, x_2) \left( z_2 \gamma^a \overleftarrow{\partial}_{2a} - \frac{(d - 1)}{2} \gamma_0 \right) = \\
&\frac{\gamma_0 \gamma_{a} \bar{x}_{12}^{a}}{ \sqrt{z_1 z_2} \sqrt{\zeta + 4} } \left( \zeta (\zeta + 4) \alpha'' (\zeta) + d (\zeta + 2) \alpha'(\zeta) + \frac{(d -1)}{4} \left( d - \frac{\zeta}{\zeta + 4} \right) \alpha(\zeta) \right) \\
&- \frac{\gamma_a x_{12}^a}{\sqrt{z_1 z_2} \sqrt{\zeta}} \left( \zeta (\zeta + 4) \beta'' (\zeta) + d (\zeta + 2) \beta'(\zeta) + \frac{(d -1)}{4} \left( d - \frac{(\zeta + 4)}{\zeta} \right) \beta(\zeta) \right) 
\end{split}
\end{equation}
The GNY equation of motion sets this to 
\begin{equation}
(\gamma\cdot \nabla_1) G_{\Psi}(x_1, x_2) (\gamma \cdot \overleftarrow{\nabla}_2) = - g_1^2 \langle s(x_1) s(x_2) \rangle G_{\Psi}(x_1, x_2). 
\end{equation}
In addition to the choice of boundary condition for the fermion, we now have an additional choice for the boundary condition on the scalar. If we choose Neumann boundary condition on the scalar, then we get the following differential equations for $\alpha$ and $\beta$ 
\begin{equation}
\begin{split}
&\zeta (\zeta + 4) \alpha'' (\zeta) + d (\zeta + 2) \alpha'(\zeta) + \frac{d -1}{4} \left( d - \frac{\zeta}{\zeta + 4} \right) \alpha(\zeta) =  \frac{\pm 2 \epsilon}{2 N + 3} \left( \frac{1}{\zeta} + \frac{1}{4 + \zeta} \right) \frac{1}{(\zeta + 4)^{\frac{3}{2}}} \\
&\zeta (\zeta + 4) \beta'' (\zeta) + d (\zeta + 2) \beta'(\zeta) + \frac{d -1}{4} \left( d - \frac{\zeta + 4}{\zeta} \right) \beta(\zeta)   =  \frac{-2 \epsilon}{2 N + 3} \left( \frac{1}{\zeta} + \frac{1}{4 + \zeta} \right) \frac{1}{\zeta^{\frac{3}{2}}}.
\end{split}
\end{equation}
There is a similar equation for when we choose Dirichlet boundary condition on the scalar, apart from the fact that the propagator on the right is $1/\zeta - 1/ (4 + \zeta)$. As we did before, we may expand the differential operator and the correlator in powers of $\epsilon$
\begin{equation}
\alpha (\zeta) = \alpha_0(\zeta) + \epsilon \alpha_1(\zeta), \hspace{0.25cm} \alpha_0(\zeta) = \pm \frac{1}{(\zeta + 4)^{\frac{3}{2}}} ; \hspace{0.5cm} \beta (\zeta) = \beta_0(\zeta) + \epsilon \beta_1(\zeta),  \hspace{0.25cm} \beta_0(\zeta) = -\frac{1}{\zeta^{\frac{3}{2}}}. 
\end{equation}
Plugging these in to the differential operators above, we can solve the differential equation to get order $\epsilon$ correction to the correlator 
\begin{equation}
\begin{split}
\beta^N_1(\zeta) &= \frac{d_1}{\zeta^{\frac{3}{2}}} + \frac{ d_2}{\zeta^{\frac{3}{2}}} \left[\log \left(1 + \zeta/4 \right) - \frac{\zeta} {4 + \zeta} \right] + \frac{1}{2 (3 + 2 N) \zeta^{\frac{3}{2}}} \left( \frac{\zeta}{4 + \zeta}  - \frac{(4 N + 5) \log (\zeta/4)}{2} \right) \\
\alpha^N_1(\zeta) &= \frac{d_3}{(4 + \zeta)^{\frac{3}{2}}} + \frac{d_4}{(4 + \zeta)^{\frac{3}{2}}} \left[ \log \left( \zeta/4 \right) - \frac{\zeta + 4}{\zeta} \right] \\
& \pm  \frac{1}{2 (3 + 2 N)} \left( \frac{1}{\zeta \sqrt{4 + \zeta}} + \frac{(4 N + 7)}{2} \frac{\log (1 + \zeta/4)}{(4 + \zeta)^{\frac{3}{2}}}  \right) 
\end{split} 
\end{equation}
for Neumann boundary condition and 
\begin{equation}
\begin{split}
\beta^D_1(\zeta) &=  \frac{d_1}{\zeta^{\frac{3}{2}}} + \frac{ d_2}{\zeta^{\frac{3}{2}}} \left[\log \left(1 + \zeta/4 \right) - \frac{\zeta} {4 + \zeta} \right] - \frac{1}{2 (3 + 2 N) \zeta^{\frac{3}{2}}} \left( \frac{\zeta}{4 + \zeta}  + \frac{(4 N + 5) \log (\zeta/4)}{2} \right)  \\
& \pm  \frac{1}{2 (3 + 2 N)} \left( \frac{1}{\zeta \sqrt{4 + \zeta}} + \frac{(4 N + 5)}{2} \frac{\log (1 + \zeta/4)}{(4 + \zeta)^{\frac{3}{2}}}  \right) 
\end{split} 
\end{equation}
for Dirichlet boundary condition. Now, let's fix the undetermined coefficients. Fixing the normalization fixes $d_1 = 0$. Then, we recall that the contribution of the $\bar{\Psi} \Psi$ operator to the two-point function in the limit $x_1 \rightarrow x_2 $ i.e. $\zeta \rightarrow 0$ should look like 
\begin{equation}
G_{\Psi} (x_1,x_2) \sim \lambda_{\bar{\Psi} \Psi}^0 + \epsilon \lambda_{\bar{\Psi} \Psi}^1 + \lambda_{\bar{\Psi} \Psi}^0 \left( \frac{\gamma_{\bar{\Psi} \Psi}}{2} -\gamma_{\Psi} \right) \log \zeta
\end{equation}
It is easy to see that the constant and $\log \zeta$ terms can only appear in $\alpha(\zeta)$ and comparing their coefficient fixes 
\begin{equation}
d_4 = \pm \left( \frac{\gamma_{\bar{\Psi} \Psi}}{2} -\gamma_{\Psi} \right) = \pm \frac{(2 N + 1)}{2 (2 N + 3)}
\end{equation}
where we used the bulk data from \cite{Fei:2016sgs}. The other two constants can be determined by imposing the boundary condition \eqref{FermionBoundaryCondition} in the limit $\zeta \rightarrow \infty$ and comparing the coefficients of $1/\zeta^2$ and $\log \zeta/ \zeta^2$ in this limit 
\begin{equation}
\begin{split}
&d_2^N = \mp d_4 - \frac{1}{2 (3 + 2 N)}, \hspace{0.5cm} d_2^D = \mp d_4 \\
&d_3^N = \mp \frac{3}{2(3 + 2 N)} , \hspace{0.5cm} d_3^D = 0.
\end{split}
\end{equation}
This gives the following two-point function for the fermion in GNY model to leading order in $\epsilon$
\begin{equation}
\begin{split} 
&G^N_{\Psi} (x_1, x_2) = - \frac{\gamma_{a} x_{12}^{a} }{\sqrt{z_1 z_2}} \frac{1}{\zeta^2} \mp \frac{\gamma_0 \gamma_{a} \bar{x}_{12}^{a} }{ \sqrt{z_1 z_2}}  \frac{1}{(\zeta + 4)^2} \\
&+ \frac{\epsilon}{3 + 2 N} \bigg[  \frac{\gamma_{a} x_{12}^{a} }{\sqrt{z_1 z_2}} \left( \frac{(3 + 2 N)}{2 \zeta (4 + \zeta)}  - (N + 1) \frac{\log \left(1 + \zeta/4 \right)}{\zeta^2}  - \frac{(4 N + 5)}{4} \frac{\log (\zeta/4)}{\zeta^{2}}   \right) \\
&\pm \frac{\gamma_0 \gamma_{a} \bar{x}_{12}^{a} }{ \sqrt{z_1 z_2}}  \left(\frac{3}{2(4 + \zeta)^{2}} + \frac{N}{\zeta (4 + \zeta)} - \frac{(2 N + 1)}{2} \frac{\log \left( \zeta/4 \right)}{(4 + \zeta)^{2}} - \frac{(4 N + 7)}{4 } \frac{\log (1 + \zeta/4)}{(4 + \zeta)^{2}} \right) \bigg] 
\end{split}
\end{equation}
for the $B_2'$ phase and 
\begin{equation}
\begin{split}
&G^D_{\Psi} (x_1, x_2) = - \frac{\gamma_{a} x_{12}^{a} }{\sqrt{z_1 z_2}} \frac{1}{\zeta^2} \mp \frac{\gamma_0 \gamma_{a} \bar{x}_{12}^{a} }{ \sqrt{z_1 z_2}}  \frac{1}{(\zeta + 4)^2}  \\
&+  \frac{\epsilon}{3 + 2 N} \bigg[\frac{\gamma_{a} x_{12}^{a} }{\sqrt{z_1 z_2}} \left( \frac{N}{ \zeta (4 + \zeta)}  - \frac{(2 N + 1)}{2}  \frac{\log \left(1 + \zeta/4 \right)}{\zeta^2}  - \frac{(4 N + 5)}{4} \frac{\log (\zeta/4)}{\zeta^{2}}   \right)  \\
&\pm \frac{\gamma_0 \gamma_{a} \bar{x}_{12}^{a} }{ \sqrt{z_1 z_2}} \left(\frac{N}{\zeta (4 + \zeta)} - \frac{(2 N + 1)}{2} \frac{\log \left( \zeta/4 \right)}{(4 + \zeta)^{2}} - \frac{(4 N + 5)}{4 } \frac{\log (1 + \zeta/4)}{(4 + \zeta)^{2}} \right) \bigg] 
\end{split}
\end{equation}
for the $B_2$ phase. At large $N$, both of them go to the large $N$ result \eqref{LargeNFermTwoPP2}. BCFT data can be extracted from the two-point function. For instance, looking at it in the limit of large $\zeta$ gives us following dimensions of the leading boundary operator in $B_2'$ and $B_2$ phase 
\begin{equation}
\hat{\Delta}^N = \frac{3}{2} - \frac{(8 N + 9)}{4 (3 + 2 N)} \epsilon, \hspace{1 cm} \hat{\Delta}^D = \frac{3}{2} - \frac{(8 N + 7)}{4 (3 + 2 N)} \epsilon.
\end{equation} 
These are also in agreement with the large $N$ result of $d - 5/2$. A curious observation is that for $N = 1/4$, the anomalous dimensions of boson and fermion agree, such that to leading order in $\epsilon$, the following relations hold
\begin{equation}
\hat{\Delta}^N = \frac{3}{2} - \frac{11 \epsilon}{14} =  \hat{\Delta}^N_s + \frac{1}{2}, \hspace{1cm} \hat{\Delta}^D = \frac{3}{2} - \frac{9 \epsilon}{14} =  \hat{\Delta}^D_s - \frac{1}{2} 
\end{equation} 
as can be checked by recalling \eqref{AnomalousOneBoxScalN} and \eqref{AnomalousOneBoxScalD}. This may be related to the observation in \cite{Fei:2016sgs} that in $d = 4  - \epsilon$, for $N = 1/4$, the GNY model respects $\mathcal{N} = 1$ emergent supersymmetry, to order $\epsilon^2$. It will be interesting to check if the boundary preserves this supersymmetry. 

It is also possible to apply the equation of motion to the fermion two-point function in the large $N$ theory. In \cite{Giombi:2020rmc}, this was used to get 1/N correction to the boundary anomalous dimension for $O(N)$ BCFT. However, this requires deriving bulk and boundary channel conformal blocks for the fermion two-point function, which we did not pursue here. We hope to come back to this question in a future work. Knowing the conformal block expansion for fermion two-point function will also be useful to extract BCFT data from the results we obtained in this section in the $\epsilon$ expansion.

\section*{Acknowledgments}
This research was supported in part by the US NSF under Grant No.~PHY-1914860.

\appendix

\section{$\sigma$ propagator} \label{AppendixSigmaProp}
To obtain the $\sigma$ propagator at large $N$ one should solve the following inversion problem \eqref{SigmaPropInversion}
\begin{equation} \label{InversionEqApp}
\int d^d x \sqrt{g} H(\zeta_{x_1, x}) G_{\sigma} (\zeta_{x, x_2}) =  \frac{1}{N}  \frac{\delta^d (x_1 - x_2) }{\sqrt{g_{x_1}}}, \hspace{0.5cm} \zeta_{x_1, x_2} = \frac{(z_1 - z_2)^2 + \textbf{x}_{12}^2}{z_1 z_2}.
\end{equation}
In our case, $H(\zeta_{x_1, x}) = \mathrm{Tr} \left[ G_{\Psi}(x_1, x) G_{\Psi}(x, x_1) \right]$. Such a problem was discussed on half space in \cite{McAvity:1995zd} and the problem is essentially identical on hyperbolic space as discussed in \cite{Giombi:2020rmc}. All the details can be found in those two papers, so we will be brief. As a first step, we can integrate over the boundary coordinates as follows 

\begin{equation}
\int d^{d-1} \textbf{x}_1 H(\zeta_{x_1 x_2}) = \frac{\pi^{\frac{d-1}{2}} \left(z_1 z_2 \right)^{\frac{d-1}{2}}}{\Gamma \left(\frac{d-1}{2} \right)} \int_0^{\infty} du u^{\frac{d-3}{2}} H \left(\rho_{z_1 z_2} + u \right) = \left(z_1 z_2 \right)^{\frac{d-1}{2}} h(\rho_{z_1 z_2})
\end{equation} 
where $\rho_{z_1 z_2} = (z_1 - z_2)^2/ z_1 z_2$. This transform can be inverted as 
\begin{equation} \label{InvBoundCoord}
H (\zeta) = \frac{1}{\pi^{\frac{d-1}{2}} \Gamma \left(\frac{-d + 1}{2} \right) } \int_0^{\infty} d \rho \rho^{\frac{- d - 1}{2}} h(\rho+ \zeta) 
\end{equation}
Applying this to \eqref{InversionEqApp} and changing variables to $z = e^{2 \theta} $ gives 
\begin{equation}
\begin{split}
\int d \theta \ h \left(4 \sinh^2 (\theta_1 - \theta) \right) g_{\sigma}  \left( 4\sinh^2 (\theta - \theta_2) \right) =  \frac{\delta(\theta_1 - \theta_2)}{4 N}. 
\end{split}
\end{equation}
This can be Fourier transformed as 
\begin{equation}
\begin{split}
\tilde{h} (k) = \int d \theta e^{i k \theta} h \left(  4 \sinh^2 \theta  \right) \implies \tilde{h} (k) \tilde{g}_{\sigma} (k) =  \frac{1}{4 N}.
\end{split}
\end{equation}

Then, following \cite{McAvity:1995zd}, consider the function 
\begin{equation}
\tilde{g}_{a,b} (k) = \frac{\Gamma \left(a - \frac{i}{4} k \right) \Gamma \left(a + \frac{i}{4} k \right) }{\Gamma \left(b - \frac{i}{4} k \right) \Gamma \left(b + \frac{i}{4} k \right)}
\end{equation}
The inverse Fourier transform of the above function gives 
\begin{equation}
\begin{split}
g_{a,b} \left( 4 \sinh^2 \theta \right) &= \frac{1}{2 \pi} \int d k e^{-i k \theta} \tilde{g}_{a,b} (k) \\
&= \frac{4 \Gamma (2 a)}{ \Gamma (b - a) \Gamma(b + a) } \frac{1}{(4 \cosh^2 \theta)^{2 a}} {}_2 F_1 \left(  2a, a + b - \frac{1}{2}; 2 a + 2 b - 1; \frac{1}{\cosh^2 \theta} \right).
\end{split}
\end{equation}
We can then transform it into a function of $\zeta$ by writing the hypergeometric as a sum and using 
\begin{equation} \label{RhoIntegralTrick}
\frac{1}{\Gamma(\lambda)} \int_0^{\infty} d u u ^{\lambda - 1} \frac{\Gamma(p + \lambda)}{(4 + \rho + u)^{p + \lambda}} = \frac{\Gamma(p)}{(4 + \rho)^p}.
\end{equation}
This gives 
\begin{equation}
G_{a,b} (\zeta) = \frac{4 \Gamma \left( 2 a + \frac{d -1}{2} \right)}{ \Gamma (b - a) \Gamma(b + a) \pi^{\frac{d -1}{2}} (\zeta)^{2 a + \frac{d -1}{2}} } {}_2 F_1 \left(  2 a + \frac{d -1}{2}, a + b - \frac{1}{2}; 2 a + 2 b - 1; - \frac{4}{\zeta} \right).
\end{equation}

For $\sigma^* = d/2 - 1$, we have 
\begin{equation}
H(\zeta) =-\frac{ 4^d \Gamma \left(\frac{d}{2} \right)^2 c_d}{16 \pi^{d} \left( \zeta (4 + \zeta)\right)^{d-1}} = \frac{ d \Gamma \left(\frac{d}{2} \right)^2 c_d \Gamma \left(-\frac{d}{2} \right)}{16 (\pi)^{\frac{d}{2}}  \Gamma(d-1)} G_{\frac{3 (d-1)}{4}, \frac{d + 1}{4}} (\zeta).
\end{equation}
This gives 
\begin{equation}
\tilde{g}_{\sigma} (k) =  \frac{4 (\pi)^{\frac{d}{2}}  \Gamma(d-1)}{d N \Gamma \left(\frac{d}{2} \right)^2 c_d \Gamma \left(-\frac{d}{2} \right)} \tilde{g}_{\frac{d + 1}{4},\frac{3 (d-1)}{4}} (k)
\end{equation}
which gives the $\sigma$ propagator 
\begin{equation} \label{SigmaCorrAppP1}
\begin{split}
G_{\sigma} (\zeta) &=   \frac{4 (\pi)^{\frac{d}{2}} \Gamma(d-1)}{d N \Gamma \left(\frac{d}{2} \right)^2 c_d \Gamma \left(-\frac{d}{2} \right)} G_{\frac{d + 1}{4},\frac{3 (d-1)}{4}} (\zeta ) \\ &=  - \frac{2^{4 d-5} (d-2) \Gamma \left(\frac{d-1}{2}\right)^2 \Gamma (d)}{ N c_d \pi  \Gamma \left(\frac{d}{2}\right) \Gamma \left(1-\frac{d}{2}\right) \Gamma (2 d-2) \zeta^d} \ {}_2 F_1 \left(d, d- 1, 2 d - 2 , -\frac{4}{\zeta} \right).
\end{split}
\end{equation}

For $\sigma^* = d/2 - 2$, we have 
\begin{equation}
\begin{split}
&H(\zeta) = -\frac{ 4^d \Gamma \left(\frac{d}{2} - 1 \right)^2 c_d}{64 \pi^{d} } \left( \frac{(d-2)^2}{\left( \zeta (4 + \zeta)\right)^{d-1}} +  \frac{(d-1)(d-3)}{4} \frac{1}{\left( \zeta (4 + \zeta)\right)^{d-2}} \right) \\
&= -\frac{\Gamma \left(\frac{d}{2} - 1 \right)^2 c_d \Gamma \left(2 - \frac{d}{2} \right) }{32 \pi^{\frac{d}{2}} \Gamma(d-2) } \left( - 2 G_{\frac{3 (d-1)}{4}, \frac{d + 1}{4}} (\zeta)  + (d-1) (d-3) G_{\frac{3d-7}{4}, \frac{d + 1}{4}} (\zeta) \right).
\end{split}
\end{equation}
This gives 
\begin{equation} \label{SigmaP2FourierSpace}
\begin{split}
&\tilde{h}(k) = \frac{\Gamma \left(\frac{d}{2} - 1 \right)^2 c_d \Gamma \left(2 - \frac{d}{2} \right) }{64 (4 \pi^{\frac{d}{2}}) \Gamma(d-2) } \left( k^2 + (d-5)^2 \right ) \tilde{g}_{\frac{3d-7}{4}, \frac{d + 1}{4}} (k) \\
\implies &\left( k^2 + (d-5)^2 \right ) \tilde{g}_{\sigma} (k) = \mathcal{B}  \pi^{\frac{d-1}{2}} \tilde{g}_{ \frac{d + 1}{4}, \frac{3d-7}{4}}(k)
\end{split}
\end{equation}
where 
\begin{equation} \label{DefnB}
\mathcal{B} = \frac{64 \pi^{\frac{1}{2}} \Gamma(d-2) }{ N \Gamma \left(\frac{d}{2} - 1 \right)^2 c_d \Gamma \left(2 - \frac{d}{2} \right)} .
\end{equation}
This gives the following differential equation for the $\sigma$ propagator in terms of $\rho$ 
\begin{equation}
\begin{split}
&\left( - \frac{d^2}{d \theta^2} + (d-5)^2  \right) g_{\sigma} (4 \sinh^2 \theta) = \mathcal{B}  \pi^{\frac{d-1}{2}} g_{ \frac{d + 1}{4}, \frac{3d-7}{4}}(4 \sinh^2 \theta) \implies \\
 & \left( \rho (4 + \rho) \frac{d^2}{d \rho^2} + (\rho + 2) \frac{d}{d \rho} -  \left( \frac{d - 5}{2} \right)^2 \right) g_{\sigma} (\rho) = -\frac{\mathcal{B}  \pi^{\frac{d-1}{2}} }{4} g_{ \frac{d + 1}{4}, \frac{3d-7}{4}}(\rho)
\end{split}
\end{equation}
We can then use \eqref{InvBoundCoord} to get a differential equation in terms of $\zeta$
\begin{equation} 
\begin{split}
&\left( \zeta (4 + \zeta) \frac{d^2}{d \zeta^2} + d(\zeta + 2) \frac{d}{d \zeta} + 2(d-3) \right) G_{\sigma} (\zeta) \\
&= \frac{2^{d+3} \sin \left(\frac{\pi  d}{2}\right) \Gamma \left(\frac{d-1}{2}\right) \Gamma (d)}{ N c_d \pi  \Gamma \left(\frac{d}{2}-2\right) \Gamma \left(d-\frac{3}{2}\right) \zeta^d} \ {}_2 F_1 \left(d, d - 2, 2 d - 4, - \frac{4}{\zeta} \right).  
\end{split}
\end{equation}
The differential equation has a solution of the form  
\begin{equation} \label{SigmaPropP2Gen}
\begin{split}
&G_{\sigma} (\zeta)  = G^P_{\sigma} (\zeta) \ + \\
& \frac{c_1}{(4 + \zeta)^2} {}_2 F_1 \left(2, 3 - \frac{d}{2}, 6 - d; \frac{4}{4 + \zeta} \right) + \frac{c_2}{(4 + \zeta)^{d-3}} {}_2 F_1 \left(d-3, \frac{d}{2} - 2, d-4; \frac{4}{4 + \zeta} \right) 
\end{split}
\end{equation}
where $G^P_{\sigma} (\zeta)$ is the particular solution and the second line is the solution to the homogeneous equation. To calculate the particular solution, we recall from \eqref{SigmaP2FourierSpace}
\begin{equation}
\tilde{g}^P_{\sigma} (k) = \frac{\mathcal{B}  \pi^{\frac{d-1}{2}} }{ \left( k^2 + (d-5)^2 \right )} \frac{\Gamma \left( \frac{d + 1}{4} - \frac{i k }{4} \right) \Gamma \left( \frac{d + 1}{4} + \frac{i k }{4} \right)}{\Gamma \left( \frac{3d-7}{4} - \frac{i k }{4} \right) \Gamma \left( \frac{3d-7}{4} + \frac{i k }{4} \right)}
\end{equation}
We then need to perform a Fourier transform of this 
\begin{equation} \label{ContourIntegral}
g^P_{\sigma} (\rho = 4 \sinh^2 \theta) = \frac{1}{2 \pi} \int d k e^{- i k \theta} \tilde{g}_{\sigma} (k).
\end{equation}
We can do the integral by a contour integration in the upper half $k-$ plane for $\theta < 0$ while in the lower half $k-$ plane for $\theta > 0$. There are poles at $\pm i (5-d)$ and $\pm i (d + 1 + 4 n)$. The arc at infinity can be dropped for $d > 3$, which is the region we are interested in
\begin{equation}
\begin{split}
g^P_{\sigma} (\rho = 4 \sinh^2 \theta) &= -\mathcal{B}  \pi^{\frac{d-1}{2}} \bigg[ \frac{ \pi  2^{3-d} e^{-((5 - d) |\theta| )}}{(d-5)^2 \Gamma \left(\frac{d-5}{2}\right) \Gamma \left(\frac{d-1}{2}\right)}\\
&+ \sum_{n = 0}^{\infty} \frac{(-1)^n e^{-  (d+4 n+1) |\theta|} \Gamma \left(\frac{d+1}{2}+n\right)}{(2 n+3) n! (d+2 n-2) \Gamma \left(\frac{d}{2}-n-2\right) \Gamma \left(d+n-\frac{3}{2}\right)}  \bigg].
\end{split}
\end{equation}
Recall that 
\begin{equation}
e^{- 2 a |\theta|} = \left(\frac{\sqrt{\rho} + \sqrt{4 + \rho}}{2} \right)^{- 2 a} = \frac{1}{(4 + \rho)^a} {}_2 F_1 \left(a, a + \frac{1}{2}, 2 a + 1, \frac{4}{4 + \rho} \right)
\end{equation}
and then using \eqref{RhoIntegralTrick}, we can do the integral over $\rho$ 
\begin{equation}
\begin{split}
&\frac{1}{\pi^{\frac{d-1}{2}} \Gamma \left(\frac{-d + 1}{2} \right) } \int_0^{\infty} d \rho \rho^{\frac{- d - 1}{2}}\frac{1}{(4 + \rho + \zeta)^a} {}_2 F_1 \left(a, a + \frac{1}{2}, 2 a + 1, \frac{4}{4 + \rho + \zeta} \right) \\
&=  \frac{\Gamma \left( a + \frac{d - 1}{2} \right)}{\pi^{\frac{d-1}{2}} \Gamma \left( a \right) } \frac{1}{(4 + \zeta)^{a + \frac{d - 1}{2}}} {}_2 F_1 \left(a + \frac{d - 1}{2}, a + \frac{1}{2}, 2 a + 1, \frac{4}{4 + \zeta}  \right) \\
&= \frac{\Gamma \left( a + \frac{d - 1}{2} \right)}{\pi^{\frac{d-1}{2}} \Gamma \left( a \right) } \frac{\zeta + 2}{(\zeta(4 + \zeta)^{ \frac{ 2 a + d + 1}{4}}} {}_2 F_1 \left(\frac{ 2 a + d + 1}{4}, \frac{ 2 a - d + 5}{4},  a + 1, -\frac{4}{\zeta(4 + \zeta)}  \right)  .
\end{split}
\end{equation}
This gives 
\begin{equation}
\begin{split}
&G^P_{\sigma} (\zeta) = -\mathcal{B} \bigg[ \frac{2^{2-d}  \cos \left(\frac{\pi  d}{2}\right)}{(d-5) \Gamma \left(\frac{d-1}{2}\right)} \frac{1}{(4 + \zeta)^{2}} {}_2 F_1 \left(2, 3 - \frac{ d}{2}, 6 - d , \frac{4}{(4 + \zeta)}  \right) \\
&+ \sum_{n = 0}^{\infty} \frac{(-1)^n  \Gamma \left(\frac{d+1}{2}+n\right) \Gamma (d+2 n) (\zeta + 2) \, _2F_1\left(\frac{d + 1}{2} +  n, n + \frac{3}{2} ; \frac{d+4 n+3}{2}; -\frac{4}{\zeta(\zeta +4)}\right)}{(2 n+3) n! (d+2 n-2) \Gamma \left(\frac{d}{2}-n-2\right) \Gamma \left(d+n-\frac{3}{2}\right) \Gamma \left(\frac{d+4 n+1}{2} \right)(\zeta(\zeta +4))^{\frac{d + 1}{2} +  n} }  \bigg].
\end{split}
\end{equation}
The first term is also a solution to the homogeneous equation, so we do not need to include it in the particular solution. So we focus on the sum in the second line. By expanding the hypergeometric, it can be rewritten as 
\begin{equation}
G_{\sigma}^P (\zeta) = -\mathcal{B} \frac{\zeta + 2}{(\zeta(4 + \zeta))^{\frac{d + 1}{2}}} \sum_{N = 0}^{\infty} \frac{h_N}{N!} \left( -\frac{4}{\zeta(4 + \zeta)} \right)^N
\end{equation}
where
\begin{equation} 
\begin{split}
h_N &= \sum_{n = 0}^N \frac{ N! 4^{-n} \Gamma \left(\frac{d+1}{2}+n\right) \Gamma (d+2 n) \left(\frac{d + 1}{2}+  n \right)_{N - n} ( n + \frac{3}{2})_{N -  n} \left(\frac{d+4 n+1}{2} \right)  (\frac{d + 3 }{2} + n + N)_{n - N} }{(2 n+3) n! (N -  n)! (d+2 n-2) \Gamma \left(\frac{d}{2}-n-2\right) \Gamma \left(d+n-\frac{3}{2}\right) \Gamma } \\
&= \frac{\Gamma(d)  \left(\frac{d + 1}{2} \right)_N \left(\frac{3}{2} \right)_N}{3 \Gamma \left(d - \frac{3}{2} \right) (d - 2)  \Gamma \left(\frac{d}{2}  - 2\right)  \left( \frac{d + 3}{2} \right)_N} {}_5 F_4 \left[ \begin{matrix}
 \frac{d + 1}{2}, \frac{d + 5}{4}, \frac{d}{2} - 1, 3 - \frac{d}{2}, - N; \\ \\
 \frac{d + 1}{4}, \frac{5}{2}, d - \frac{3}{2}, \frac{d + 3}{2} + N; 
\end{matrix} \ \ 1  \right].
\end{split}
\end{equation}
Using a special case of Dougall's theorem \cite{SlaterHypergeometric}, we get \begin{equation}
h_N = \frac{\Gamma(d)  \left(\frac{d + 1}{2} \right)_N \left(\frac{3}{2} \right)_N \left( \frac{d - 1}{2} \right)_N}{3 \Gamma \left(d - \frac{3}{2} \right) (d - 2)  \Gamma \left(\frac{d}{2}  - 2\right)  \left( d - \frac{3}{2} \right)_N \left( \frac{5}{2} \right)_N}.
\end{equation} 
This finally determines the particular solution
\begin{equation}
\begin{split}
&G_{\sigma}^P (\zeta) = \\
&\frac{-\mathcal{B}\Gamma(d)}{3 \Gamma \left(d - \frac{3}{2} \right) (d - 2)  \Gamma \left(\frac{d}{2}  - 2\right)  } \frac{\zeta + 2}{(\zeta(4 + \zeta))^{\frac{d + 1}{2}}} {}_3 F_2 \left( \frac{d + 1}{2}, \frac{d - 1}{2}, \ \frac{3}{2}; d - \frac{3}{2},  \frac{5}{2};  -\frac{4}{\zeta(4 + \zeta)}\right).
\end{split}
\end{equation}
This equation, along with \eqref{SigmaPropP2Gen} gives us a general solution for the $\sigma$ correlator in this phase. To fix the constants, we note that at the boundary of hyperbolic space, $\zeta \rightarrow \infty$, there are two possible decays: $\zeta^{-2}$ or $\zeta^{3 - d}$, and they correspond to having a scalar of dimension $2$ or $d - 3$ in the boundary spectrum, respectively. For the former case, we set $c_2 = 0$.  To fix $c_1$, we look at the bulk limit of the correlator, $\zeta \rightarrow 0$. In this limit, we expect the leading term to come from identity operator in the bulk channel, and hence should fall off as $\zeta^{-1}$ since the $\sigma$ operator in the bulk has dimension $1$ at large $N$. This fixes $c_1$ and hence the correlator
\begin{equation}\label{SigmaCorrAppP21}
\begin{split}
&G^D_{\sigma} (\zeta) = -\mathcal{B} \bigg[ \frac{2^{2-d}  \cos \left(\frac{\pi  d}{2}\right)}{(d-5) \Gamma \left(\frac{d-1}{2}\right)} \frac{1}{(4 + \zeta)^{2}} {}_2 F_1 \left(2, 3 - \frac{ d}{2}, 6 - d , \frac{4}{(4 + \zeta)}  \right) \\
&+ \frac{\Gamma(d)}{3 \Gamma \left(d - \frac{3}{2} \right) (d - 2)  \Gamma \left(\frac{d}{2}  - 2\right)  } \frac{\zeta + 2}{(\zeta(4 + \zeta))^{\frac{d + 1}{2}}} {}_3 F_2 \left( \frac{d + 1}{2}, \frac{d - 1}{2}, \frac{3}{2}; d - \frac{3}{2},  \frac{5}{2};  -\frac{4}{\zeta(4 + \zeta)}\right) \bigg].
\end{split}
\end{equation}
If we instead demand that the propagator falls of as $\zeta^{3-d}$ at the boundary, we set $c_1 = 0$. The same argument as above fixes $c_2$ and the correlator turns out to be
\begin{equation}\label{SigmaCorrAppP22}
\begin{split}
&G^N_{\sigma} (\zeta) = -\mathcal{B} \bigg[ -\frac{\pi ^{\frac{1}{2}} \Gamma \left(\frac{d}{2}-1\right)}{8 \Gamma \left(\frac{d-3}{2}\right) \Gamma \left(\frac{d-1}{2}\right)} \frac{1}{(4 + \zeta)^{d-3}} {}_2 F_1 \left(d-3, \frac{d}{2} - 2, d - 4 , \frac{4}{(4 + \zeta)}  \right) \\
&+ \frac{\Gamma(d)}{3 \Gamma \left(d - \frac{3}{2} \right) (d - 2)  \Gamma \left(\frac{d}{2}  - 2\right)  } \frac{\zeta + 2}{(\zeta(4 + \zeta))^{\frac{d + 1}{2}}} {}_3 F_2 \left( \frac{d + 1}{2}, \frac{d - 1}{2}, \frac{3}{2}; d - \frac{3}{2},  \frac{5}{2};  -\frac{4}{\zeta(4 + \zeta)}\right) \bigg].
\end{split}
\end{equation}

\bibliographystyle{ssg}
\bibliography{FermionicBCFT-bib}

\end{document}